\documentclass[12pt]{article}
\usepackage{graphicx,psfrag,epsf}
\usepackage{enumerate}
\usepackage{natbib}
\usepackage{url} 

\usepackage[a4paper]{geometry}
\usepackage[utf8]{inputenc}
\usepackage{lipsum}

\usepackage{amsmath,amssymb}
\usepackage[colorlinks,citecolor=blue,urlcolor=blue]{hyperref}

\usepackage{lmodern}
\usepackage{anyfontsize}
\usepackage{dsfont}
\usepackage{epstopdf}
\usepackage[hang]{subfigure}
\usepackage{booktabs}
\usepackage{textcomp}
\usepackage{ulem}
\usepackage{color}
\usepackage{threeparttable}
\usepackage{comment}
\usepackage{float}
\usepackage{indentfirst}

\newtheorem{theorem}{Theorem}

\newtheorem{prop}{Proposition}

\newtheorem{condition}{Condition}
\newtheorem{remark}{Remark}
\newtheorem{corollary}{Corollary}

\def\T{{ \mathrm{\scriptscriptstyle T} }}
\def\t{ \mathrm{\scriptscriptstyle T} }

\def\X{\boldsymbol X}
\def\V{\boldsymbol V}
\def\S{\boldsymbol S}
\def\bbeta{\boldsymbol \beta}
\def\bgamma{\boldsymbol \gamma}
\def\bGamma{\boldsymbol \Gamma}
\def\btau{\boldsymbol \tau}
\def\s{\boldsymbol s}
\def\Y{\boldsymbol Y}

\def\dq{\boldsymbol{d}_q}
\def\dqq{\boldsymbol{d}_{q'}}
\def\R{\boldsymbol R}
\def\be{\boldsymbol \varepsilon}
\def\bmu{\boldsymbol \mu}

\def\obs{\textnormal{obs}}
\def\unadj{\textnormal{unadj}}
\def\adj{\textnormal{adj}}
\def\cond{\textnormal{cond}}
\def\condfull{\textnormal{cond2}}
\def\adjtwo{\textnormal{inter}}
\def\inter{\textnormal{inter}}
\def\var{\textnormal{var}}

\def\sumq{\sum_{q=1}^{Q}}
\def\sumi{\sum_{i=1}^{n}}
\def\summ{\sum_{m=1}^{M}}
\def\sumim{\sum_{i \in [m]}}

\def\bx{ \bar{\X} }

\def\brq{ \bar{R}(q) }
\def\brmq{\bar R_{[m]}(q)}
\def\bxmq{\bar \X_{[m]}(q)}
\def\bxm{\bar \X_{[m]}}

\def\hbym{ \widehat{ \bar{ Y }}_{[m]} }

\def\hbyq{\widehat{\bar{Y}}(q)}
\def\hbxq{\widehat{\bar{\X}}(q)}
\def\hbrq{\widehat{\bar{R}}(q)}

\def\hbymq{\widehat{ \bar Y}_{[m]}(q)}
\def\hbxmq{\widehat{\bar \X}_{[m]}(q)}
\def\hbrmq{\widehat{\bar R}_{[m]}(q)}

\def\nm{n_{[m]}}
\def\nmq{n_{[m]q}}

\def\emq{e_{[m]q}}

\def\taumk{\tau_{[m]k}}
\def\taumf{\tau_{[m]f}}

\def\pim{\pi_{[m]}}

\def\e{\varepsilon}
\def\argmin{\mathop{\arg\min}}
\def\err{\eta}
\def\berr{\boldsymbol \eta}

\def\cov{\textnormal{cov}}
\def\var{\textnormal{var}}

\def\pr{P}

\newcommand\iotaq[1]{\iota_{q,#1}}
\newcommand\iotak[1]{\iota_{#1,k}}
\newcommand\taum[1]{\tau_{[m]#1}}

\newcommand\blue[1]{\textcolor{black}{#1}}
\newcommand\red[1]{\textcolor{black}{#1}}

\newcommand{\blind}{0}

\begin{document}

\def\spacingset#1{\renewcommand{\baselinestretch}%
{#1}\small\normalsize} \spacingset{1}


\if0\blind
{
  \title{\bf Randomization-based joint central limit theorem and efficient covariate adjustment in randomized block $2^K$ factorial experiments}
  \author{Hanzhong Liu\thanks{
    Dr. Hanzhong Liu was supported by the \textit{National Natural Science Foundation of China (Grant No. 12071242)} and the Guo Qiang Institute of Tsinghua University.}, \   Jiyang Ren \hspace{.2cm}\\
    Center for Statistical Science, Department of Industrial Engineering, \\Tsinghua University, Beijing, 100084, China \vspace{.2cm}\\
    Yuehan Yang\thanks{Corresponding author: yyh@cufe.edu.cn. Dr. Yuehan Yang was supported by the \textit{National Natural Science Foundation of China (Grant No. 12001557)}, the Youth Talent Development Support Program (QYP202104),  the Emerging Interdisciplinary Project, and the Disciplinary Funding of Central University of Finance and Economics.}\hspace{.2cm}\\
    School of Statistics and Mathematics,\\ Central University of Finance and Economics, Beijing, 102206, China.}
    \date{}
  \maketitle
} \fi

\if1\blind
{
  \bigskip
  \bigskip
  \bigskip
  \begin{center}
    {\LARGE\bf Randomization-based joint central limit theorem and efficient covariate adjustment in randomized block $2^K$ factorial experiments}
\end{center}
  \medskip
} \fi

\bigskip
\begin{abstract}
Randomized block factorial experiments are widely used in industrial engineering, clinical trials, and social science. Researchers often use a linear model and analysis of covariance to analyze experimental results; however, limited studies have addressed the validity and robustness of the resulting inferences because assumptions for a linear model might not be justified by randomization in randomized block factorial experiments. In this paper, we establish a new finite population joint central limit theorem for usual (unadjusted) factorial effect estimators in randomized block $2^K$ factorial experiments. Our theorem is obtained under a randomization-based inference framework, making use of an extension of the vector form of the Wald--Wolfowitz--Hoeffding theorem for a linear rank statistic. It is robust to model misspecification, numbers of blocks, block sizes, and propensity scores across blocks. To improve the estimation and inference efficiency, we propose four covariate adjustment methods. We show that under mild conditions, the resulting covariate-adjusted factorial effect estimators are consistent, jointly asymptotically normal, and generally more efficient than the unadjusted estimator. In addition, we propose Neyman-type conservative estimators for the asymptotic covariances to facilitate valid inferences. Simulation studies and a clinical trial data analysis demonstrate the benefits of the covariate adjustment methods.
\end{abstract}

\noindent%
\vfill

\newpage
\spacingset{1.45} 

\section{Introduction}

Since initially proposed by \citet{Fisher1935} and  \citet{Yates1937}, factorial experiments have been widely used to study the joint effects of several factors on a response \citep[see, e.g.,][]{Cochran1950,Angrist2009,Wu2011,Dasgupta2015}. Consider a $2^K$ factorial experiment with $ n $ units and $ K $ factors  ($K \geq 1$). Each factor has two levels, $ -1 $ and $ +1 $, and there are $ Q = 2^K $ treatment combinations. Complete randomization of the treatment combinations can balance the covariates on average; however, as the numbers of baseline covariates and factors increase, it is likely that some covariates will exhibit imbalance in a particular treatment assignment, as observed in completely randomized experiments \citep{Fisher1926, Senn1989, Morgan2012}. Blocking, or stratification, which was initially proposed by \cite{Fisher1926} and suggested by classical experimental design textbooks \citep[e.g.,][]{box2005, Wu2011}, is the most common way to balance treatment allocations with respect to a few discrete variables that are most relevant to the response. Appropriate blocking can balance the baseline covariates and improve the treatment effect estimation efficiency \citep[e.g.,][]{wilk1955randomization,imai2008variance,Imbens2015}. A recent survey  \citep{Lin2015} noted that 70\% of the 224 randomized trials published in leading medical journals in 2014 had used blocking (or stratification) in the experimental design. Even when blocking (or stratification) is not used in the design stage, researchers have recommended its use at the analysis stage and have shown that this post-stratification strategy can also improve estimation efficiency \citep{mchugh1983post,miratrix2013}.

Blocking or post-stratification balances only a few discrete variables; however, in the present era of big data, researchers  often observe many other baseline covariates that are relevant to the response and that might still be imbalanced \citep{Rosenberger2008, Liu2019, Wang2021}. For example, in clinical trials, demographic and disease characteristics are often collected for each patient, and it is impossible to completely balance these covariates using only blocking. Covariate adjustment or regression adjustment is a common strategy to adjust for the remaining imbalances in the additional covariates, following a similar concept in survey sampling literature \citep[e.g.,][]{Cassel1976, Sarndal2003}. In practice,  researchers often use analysis of covariance to analyze the results of randomized block factorial experiments \citep[see textbooks][]{Cochran1950,Montgomery2013}, assuming a linear model with fixed or random block effects. However, concerns have been raised regarding the validity of the resulting inferences because the ``usual" assumptions for a linear model, such as linearity, normality, and homoskedastic errors, might not be justified by randomization in randomized block factorial experiments.


Randomization-based inference is receiving increasing attention in the field of causal inference. This inference framework allows the analysis model to be arbitrarily misspecified, and thus is more robust compared to the studies requiring a true linear model.   \cite{Li2016} established general forms of finite population central limit theorems (CLTs) to draw causal inferences in completely randomized experiments. In completely randomized $2^K$ factorial experiments, \cite{Dasgupta2015} defined factorial effects using potential outcomes and explored Fisher's randomization tests on sharp null hypotheses; and \cite{lu2016randomization} proposed valid randomization-based  inferences for the average factorial effects. However, none of these studies considered blocking used in the design stage.

Our first contribution is to establish an asymptotic theory on the joint sampling distribution of the usual (unadjusted) factorial effect estimators in randomized block $2^K$ factorial experiments, under randomization-based inference framework, without imposing strong modeling assumptions on the true data generation process. Most relevant to our work, \cite{Liu2019} used the results of \cite{bickel1984asymptotic} to establish the asymptotic normality of the blocked difference-in-means estimator in randomized block  experiments, but with one-dimensional potential outcomes and two treatments. As multiple factorial effects are simultaneously of interest in randomized block factorial experiments, it is important to determine
the joint asymptotic distribution to handle multiple treatments. In the literature, \cite{Li2016} established the joint asymptotic normality of the usual average treatment effect estimator in completely randomized experiments with multiple treatments. Their result can be easily extended to randomized block $2^K$ factorial experiments in which the number of blocks is fixed with their sizes tending to infinity. However, in many applications of factorial experiments in clinical trials and industrial engineering, the number of blocks often tends to infinity with their sizes being fixed. It is unclear whether the joint asymptotic normality of \cite{Li2016} holds in such cases. To fill in this gap, we establish the CLT of the blocked difference-in-means estimator in randomized block experiments with vector potential outcomes and multiple treatments, by making use of  the techniques for obtaining the vector-form of the Wald--Wolfowitz--Hoeffding theorem for a linear (or bi-linear) rank statistic \citep{Hajek::1961, sen1995hajek}. Our new CLT is robust to model misspecification, numbers of blocks, block sizes, and propensity scores (i.e., the proportion of units under each treatment arm in each block).

Covariate adjustment or regression adjustment is widely used in randomized experiments to balance baseline covariates and improve estimation efficiency. Recently, the asymptotic properties of covariate adjustment have been investigated under randomization-based inference framework for various experimental designs, including completely randomized experiments with two treatments \citep{Freedman2008,freedman2008randomization,lin2013, miratrix2013,bloniarz2015lasso,Lei2020}, completely randomized factorial experiments \citep{lu2016covariate, lu2016randomization}, and randomized block  experiments with two treatments
 \citep{Liu2019}. Among them, \cite{lu2016covariate} proposed a covariate adjustment method  and studied its efficiency gain in completely randomized factorial experiments. Again, this method can be directly extended to randomized block $2^K$ factorial experiments when the number of blocks is fixed with their sizes tending to infinity, but it is not applicable for general scenarios in which the number of blocks and their sizes both tend to infinity.  \citet{Liu2019} proposed a  regression adjustment method in randomized block experiments with two treatments. This method does not require the block sizes to tend to infinity, but it works only for the cases of two treatments and equal propensity scores across blocks.

Our second contribution is to propose \blue{four} covariate adjustment methods to improve the estimation efficiency of factorial effects in randomized block $2^K$ factorial experiments. The first method extends the method proposed in \cite{Liu2019} to handle multiple treatments; the second \blue{and third methods are developed from a conditional inference perspective}, which overcome the drawback of the first method regarding the requirement of equal propensity scores across blocks; and the \blue{last method is applicable in cases with only large blocks}. Under appropriate conditions, we show that the resulting covariate-adjusted factorial effect estimators are all consistent and jointly asymptotically normal. Our analysis is conducted under randomization-based inference framework,  so our results are robust to model misspecification. Moreover, we compare the efficiency of various factorial effect estimators, and show that the asymptotic covariance of the first covariate-adjusted factorial effect estimator is no greater than that of the unadjusted estimator when the propensity scores are the same across blocks; the second \blue{and third covariate-adjusted methods improve} the efficiency even when the propensity scores differ across blocks; and the \blue{last method is generally more efficient than the first three}, but it requires the number of blocks to be fixed, with their sizes tending to infinity. In addition, we propose conservative estimators for the asymptotic covariances that can be used to construct large-sample conservative confidence intervals or regions for the factorial effects.

The paper proceeds as follows. Section 2 introduces the framework and notation of randomized block $2^K$ factorial experiments. Section 3 establishes the joint CLT for the unadjusted factorial effects estimator. Section 4 proposes four covariate adjustment methods to improve estimation efficiency and studies their asymptotic properties. Section 5 provides an extensive simulation study. Section 6 contains an application to a clinical trial dataset. Section 7 concludes the paper with discussions. The proofs are given in the Supplementary Material.


\section{Framework and notation}

We follow the framework and notation introduced in \cite{Dasgupta2015} and \cite{Li2020factorial} for $2^K$ factorial experiments and generalize them to the situation in which the design stage uses blocking.

\subsection{Potential outcomes and average factorial effects}

In a randomized block $2^K$ factorial experiment with $ n $ units and $ K $ factors, each factor has two levels, $ -1 $ and $ +1 $, and there are $ Q = 2^K $ treatment combinations. Before randomization, the units are blocked into $M$ blocks according to the values of some important discrete variables, such as gender, disease stage, or location. We use the subscript ``$ [m] $'' for block, ``$ k $'' for factor and ``$ i $'' for unit. The block $ m $ contains $ \nm $ units, $ \nm \geqslant Q $ and $ \sum^M_{m = 1} \nm = n$.  Within block $m$ ($m=1,\dots,M$), $\nmq $ ($\nmq \geq 1$) units are randomly assigned to the treatment combination $ q $ $(q = 1,\dots,Q)$. Thus, the total number of units under treatment combination $ q $ is $ n_q = \sum^M_{m = 1} \nmq $.  Let $Z_i $ be the treatment  assignment indicator for unit $ i $. As the treatment assignments are independent across blocks, the probability that $ \boldsymbol Z = (Z_1,\dots,Z_n)^\T $ takes a particular value $ (z_1,\dots,z_n)^\T $ is
$$\pr ( \boldsymbol Z =  \boldsymbol z ) =\prod_{m=1}^M \bigg(  \frac{ \prod_{q=1}^Q n_{[m]q}! }{ n_{[m]}! }  \bigg) , \quad \sum_{i \in [m]}I(z_i = q) = \nmq ,$$
where $ I(\cdot) $ is an indicator function, and $i \in [m]$ indexes the unit $i$ in block $m$.

We define factorial effects using the potential outcomes framework \citep{Rubin:1974,Neyman:1923}. Let us denote $Y_i(q) $ as the potential outcome of unit $i$ under treatment combination $ q $, and all $Q$ potential outcomes are denoted as a column vector $ \boldsymbol Y_i = (Y_i(1),\dots,Y_i(Q))^\T$. The unit-level factorial effects can be defined as contrasts of these potential outcomes. As each unit is assigned to only one treatment combination, we observe only one of the $Q$ potential outcomes. Therefore, the unit-level factorial effects are not identifiable without additional modeling assumptions. Fortunately, under the stable unit treatment value assumption \citep{Rubin:1980}, the average factorial effects across all experimental units are estimable. For treatment combination $q$, let $ \boldsymbol \iota_q = (\iotaq{1},\dots,\iotaq{K})^\T \in \{ -1,+1 \}^K $ be the levels of the $ K $ factors, and let $\bar{Y}_{[m]} (q) = (1/\nm) \sum_{i \in [m]} Y_i(q)$ be the mean of the potential outcome $Y_i(q)$ within block $m$. The block-specific average main effect of factor $ k $ within block $ m $ can be defined as
\begin{align*}
\taumk & = \dfrac{1}{\nm} \cdot \dfrac{2}{Q} \sum_{i \in [m]} \sum^Q_{q = 1} I(\iotaq{k} = 1)Y_i(q) -  \dfrac{1}{\nm} \cdot \dfrac{2}{Q}\sum_{i \in [m]} \sum^Q_{q = 1} I(\iotaq{k} = -1)Y_i(q)\\
& =  \dfrac{1}{2^{K-1}} \sum^Q_{q = 1} \iotaq{k} \bar{Y}_{[m]}(q)  = \dfrac{1}{2^{K-1}}   \boldsymbol  g_k^\T \bar{\Y }_{[m]} ,
\end{align*}
where  $\boldsymbol g_k = (g_{k,1},\dots,g_{k,Q})^\t = (\iotak{1},\dots,\iotak{Q})^\t $ is called the generating vector for the main effect of factor $ k $, and $  \bar{ \boldsymbol Y }_{[m]}  = (  \bar{Y}_{[m]} (1), \dots, \bar{Y}_{[m]}(Q) )^\T$. The relationship between $ \boldsymbol  \iota $ and $ \boldsymbol  g $ can be found in Table~\ref{table eg} for a  $2^3$ factorial experiment.

\begin{table}
\centering
\caption{\label{table eg}Relation of $ \boldsymbol  \iota $ and $ \boldsymbol  g $ for a $2^3$ factorial experiment}
\begin{tabular}{ccc|cccccccccc}
\hline
&& & $  \boldsymbol  \iota_1$ & $ \boldsymbol  \iota_2$ & $ \boldsymbol  \iota_3$ & $  \boldsymbol  \iota_4$
& $ \boldsymbol  \iota_5$ & $ \boldsymbol  \iota_6$ & $ \boldsymbol  \iota_7$ & $ \boldsymbol  \iota_8$ && \\ \hline
&&$ \boldsymbol  g_1$ & +1 & +1 & +1 & $ + $ 1 & $-$1 & $-1$ & $-1$ & $-1$ && \\
&&$ \boldsymbol  g_2$ & +1 & +1 & $-1$ & $-1$ & $+1$ & +1 & $-1$ & $-1$ && \\
&&$ \boldsymbol  g_3$ & +1 & $-1$ & $+1$ & $-1$ & +1 & $-1$ & $+1$ & $-1$  && \\
\hline
\end{tabular}
\end{table}

The average main effect of factor $ k $ can be defined as
$$
\tau_k
 = \dfrac{1}{n} \cdot \dfrac{2}{Q} \sum^n_{i=1} \sum^Q_{q = 1}\iotaq{k} Y_i(q)
=  \sum^M_{m=1} \pi_{[m]} \taumk,
$$
where $ \pi_{[m]} = \nm/n $ is the \blue{fraction of units in block $m$}. As shown by \cite{Dasgupta2015}, the interaction effect among several factors can be defined with the $ \boldsymbol  g $-vector, which is an element-wise multiplication of the generating vectors for the corresponding factors' main effects. More specifically, for $ 1 \leqslant f \leqslant F =   2^K - 1 = Q - 1$, let $\boldsymbol g_f = (g_{f,1},\dots,g_{f,Q})^\t \in \{-1,+1 \}^Q $ be the generating vector for the $ f $th factorial effect,  which satisfies $\sumq g_{f,q} = 0$. The $ f $th block-specific average factorial effect in block $ m $ can be defined as
$\taumf =   2^{-(K-1)}  \boldsymbol  g_f ^\T  \bar{ \boldsymbol Y }_{[m]}  .$
We denote all block-specific average factorial effects in block $m$ by an $F$-dimensional column vector $ \boldsymbol \tau_{[m]} = (\taum{1},\dots,\taum{F})^\t $. Let $\boldsymbol  d_q = (g_{1,q},\dots,g_{F,q})^\T$, $q=1,\dots,Q$, then
\[ \boldsymbol \tau_{[m]}  = \dfrac{1}{2^{K-1}}   \sum^Q_{q = 1} \dq  \bar{Y}_{[m]}(q), \quad  \sumq \dq  = \boldsymbol  0.  \]
Thus, each block-specific average factorial effect $\tau_{[m]f}$ is a linear contrast of the block-specific average potential outcomes $ \bar{Y}_{[m]}(q)$. Let us denote the average potential outcomes as
$
 \bar{Y}(q) = n^{-1} \sumi Y_i(q) =  \summ \pim \bar{Y}_{[m]}(q).
$
The vector of all average factorial effects is defined  as
\[\boldsymbol \tau = \dfrac{1}{2^{K-1}}   \sum^Q_{q = 1} \dq  \bar{Y}(q) = \sum^M_{m=1} \pi_{[m]} \boldsymbol \tau_{[m]}.  \]

\blue{
There are estimands of interest such that $\boldsymbol g_f \notin \{-1, +1\}^Q$. For example, in conjoint experiments  (a specific type of factorial experiments), researchers may want to weight treatment combinations by relative prevalence in the population \citep{Cuesta2019} in place of the uniform weighting implied by $\boldsymbol g_f \in \{-1,+1\}^Q$. Such estimands can be represented by linear transformations of the average factorial effects $\boldsymbol C \boldsymbol \tau$, where $\boldsymbol C \in \mathbb{R}^{F_1 \times F}$ ($F_1 \leq F$) is a constant matrix and has full row rank.
}

Before performing physical randomization, the experimenter collects an additional $p$-dimensional vector of baseline covariates $\X_i = (X_{i1}, \dots, $ $X_{ip})^\T \in \mathbb{R}^p$ for each unit $i$. In this paper, we consider a finite population randomization-based inference, in which both the potential outcomes $\boldsymbol Y_i$ and covariates $\X_i$ are fixed quantities, and the only source of randomness is the treatment assignment $\boldsymbol Z$. Under the stable unit treatment value assumption, the observed outcome $Y_i^\obs$ is a function of the treatment assignment indicator and potential outcomes:
$  Y_i^\obs = \sumq I(Z_i = q)  Y_i(q).$
We aim to make robust and efficient inferences of the average factorial effects $\boldsymbol \tau $ \blue{(and $\boldsymbol C \boldsymbol \tau$)} using the observed data $\{Y_i^\obs, \X_i , Z_i\}_{i=1}^{n}$.

\subsection{Notation}

For potential outcomes or their transformations
$\R(q) = (R_1(q)^\T, \dots, R_n(q)^\T )^\T, \ q = 1,\dots, Q,$
where $R_i(q)$ can be a column vector, we add a bar on top and a subscript $[m]$ ($m=1,\dots,M$) to denote its block-specific mean,
$ \brmq  = (1 / \nm ) \sum_{i \in [m]} R_i(q). $
We add an additional hat to denote the corresponding block-specific sample mean,
$ \hbrmq = ( 1 / \nmq ) \sum_{i \in [m]}  I(Z_i = q) R_i(q).$
As covariates can be considered as potential outcomes that are not affected by treatment assignment, we denote $\X_i(q) = \X_i$ and $\bxmq = \bar{\boldsymbol X}_{[m]} = (1/\nm) \sum_{i \in [m]} \boldsymbol X_i.$
The overall mean is denoted as
$ \brq  = (1 / n ) \sum^n_{i=1} R_i(q) = \sum^M_{m=1}\pi_{[m]}  \brmq ,$
and  its natural unbiased estimator is denoted as
$ \hbrq =  \sum^M_{m=1}\pi_{[m]}  \hbrmq. $
For finite population quantities $\boldsymbol H = (H_1^\T,\dots, H_n^\T)^\t  $ and  $\boldsymbol U = (U_1^\T,\dots, U_n^\T)^\t$, we  denote the block-specific covariance of $\boldsymbol H  $ as $\boldsymbol \S^2_{[m] \boldsymbol H} = \S_{[m] \boldsymbol H \boldsymbol H} =   (\nm - 1)^{-1} \sum_{i \in [m]} ( H_i - \bar{H}_{[m]} ) ( H_i - \bar{H}_{[m]}  )^\T,$
and the block-specific covariance between  $\boldsymbol H $ and  $\boldsymbol U   $ as $\boldsymbol \S_{[m] \boldsymbol H \boldsymbol U} = ( \nm - 1 )^{-1} \sum_{i \in [m]} ( H_i - \bar{H}_{[m]} ) ( U_i - \bar{U}_{[m]}  )^\T.$
Here and in what follows,  both $H_i$ and $ U_i$ can be column vectors, and when they are one-dimensional real numbers, we use $S$ to replace $\boldsymbol S$. The corresponding sample quantities are denoted by  $\boldsymbol s^2_{[m] \boldsymbol H} = \boldsymbol s_{[m] \boldsymbol H \boldsymbol H}$ and $\boldsymbol s_{[m]\boldsymbol H \boldsymbol U}  $. All the above defined quantities depend on $n$, but we do not index them with $ n $ for notational simplicity. For an $L$-dimensional vector $\boldsymbol u = (u_1,\dots, u_L)^\T$, let $\| \boldsymbol u \|_1$,  $\| \boldsymbol u \|_2$ and $\|  \boldsymbol u \|_\infty$ be the $\ell_1$, $\ell_2$ and $\ell_\infty$ norms, respectively. For two matrices $A$ and $B$, we write $A \geq B$ if $A - B$ is positive semidefinite, and $A > B$ if $A-B$ is positive definite. We denote $a_n \asymp   b_n$ if $a_n$ and $b_n$ have the same order asymptotically, i.e., both the superior limits of $a_n / b_n$ and $b_n / a_n$ are bounded by constants.


\subsection{Blocked difference-in-means estimator}
The block-specific average factorial effects $ \boldsymbol \tau_{[m]}$ can be estimated without bias using a plug-in estimator $ \widehat {\boldsymbol \tau}_{[m]}  $,  which  replaces the block-specific mean $\bar{ \boldsymbol  Y}_{[m]} $ by the corresponding sample mean
$ \widehat { \bar{ \boldsymbol  Y}}_{[m]} =  ( \widehat{ \bar{Y}}_{[m]} (1), \dots, \widehat {\bar{Y}}_{[m]}(Q) )^\T;$
that is,
$  \widehat {\boldsymbol \tau}_{[m]}  =  {2^{-(K-1)}}   \sum^Q_{q = 1} \dq  \widehat{\bar{Y}}_{[m]}(q). $
Thus, an unbiased estimator for the overall average factorial effects $ \boldsymbol \tau$ is
\[ \widehat { \boldsymbol \tau }_{\unadj} = \summ  \pi_{[m]}  \widehat {\boldsymbol \tau}_{[m]} =  \dfrac{1}{2^{K-1}}   \sum^Q_{q = 1} \dq  \widehat{\bar{Y}}(q) ,   \]
where the subscript ``unadj" indicates that this estimator does not adjust for covariate imbalances. We call it the blocked difference-in-means estimator.


Let $ \emq = \nmq/\nm $ be the propensity score under treatment combination $ q $ in block $ m $. The covariance of $ \widehat { \boldsymbol \tau }_{\unadj}$ and the probability limit of its conservative estimator depend on the following $ F \times F $ covariance matrices related to $\Y$:
\[ \widetilde { \boldsymbol V }_n (\Y) =  2^{-2(K-1)} \sum^M_{m=1} \pi_{[m]}  \Big \{  \sumq \dfrac{1}{\emq} S^2_{[m]\Y(q)}  \dq \dq^\t \Big \} -  \sum^M_{m=1} \pi_{[m]}  \boldsymbol \S^2_{[m]  \boldsymbol \tau} ,\]
\[ \boldsymbol V_n (\Y) =  2^{-2(K-1)} \sum^M_{m=1} \pim  \Big \{  \sumq \dfrac{1}{\emq} S^2_{[m]\Y(q)}  \dq \dq^\t  \Big\} ,\]
where  $\boldsymbol S^2_{[m] \boldsymbol \tau}$ is the block-specific covariance of  unit-level factorial effects
$ \boldsymbol \tau_i  = {2^{-(K-1)}}   \sum^Q_{q = 1} \dq  Y_i(q);$
that is,
\[ \boldsymbol S^2_{[m] \boldsymbol \tau} = 2^{-2(K-1)} \Big \{ \sum^Q_{q = 1}  S^2_{[m]\Y(q)}  \dq \dq^\t  +  \sum_{1 \leq q \neq q' \leq Q}  S_{[m]\Y(q)\Y(q')} \dq  \dqq^\t  \Big \}. \]
Because $\sum^M_{m=1} \pi_{[m]}  \boldsymbol \S^2_{[m]  \boldsymbol \tau} \geq 0 $, it holds that $  \boldsymbol V_n (\Y)  \geq  \widetilde { \boldsymbol V }_n (\Y) $. We have the following proposition.

\begin{prop}\label{prop::mean-var}
The mean and covariance of $ \widehat{\boldsymbol \tau}_\unadj $ are
\[  E(\widehat{\boldsymbol \tau}_\unadj) =  \boldsymbol   \tau, \quad \cov( \widehat{\boldsymbol \tau}_\unadj) =  \frac{1}{n} \widetilde { \boldsymbol V}_n ( \Y ) . \]
\end{prop}

In general, $ \boldsymbol S^2_{[m]  \boldsymbol \tau} $ is not estimable because we cannot observe $ \boldsymbol \tau_i$ for any unit $i$. Fortunately, we can estimate the covariance of $ \sqrt{n} \widehat{\boldsymbol \tau}_\unadj $ using a Neyman-type conservative estimator when $\nmq \geq 2$,
\[  \widehat { \boldsymbol V}_n (\Y)  =  2^{-2(K-1)} \sum^M_{m=1} \pi_{[m]}  \Big \{ \sumq \dfrac{1}{\emq} s^2_{[m]\Y(q)}  \dq \dq^\t  \Big \}  ,\]
where $s^2_{[m]\Y(q)}$ is the sample variance of $Y_i(q)$ in block $m$,
\[  s^2_{[m]\Y(q)}  = \dfrac{1}{\nmq - 1} \sum_{i \in [m]} I(Z_i = q) \big \{ Y_i(q) -  \hbym (q) \big \}^2 .  \]
Under appropriate conditions, $ \widehat { \boldsymbol V }_n (\Y)$ converges in probability to the limit of $ {\boldsymbol V }_n (\Y)$ (see the following Theorem~\ref{thm::unadj}),
which is consistent if the unit-level factorial effects are additive in a block-specific manner, that is, $ \boldsymbol \tau_i =  \boldsymbol c_m $ for all $ i \in [m] $, where $ \boldsymbol  c_m $ is a constant vector.

To infer $ \boldsymbol \tau$ (or $ \boldsymbol C \boldsymbol \tau$)  based on $ \widehat { \boldsymbol \tau}_{\unadj} $ (or $ \boldsymbol C \widehat { \boldsymbol \tau}_{\unadj}$), we must study its joint asymptotic sampling distribution under randomization-based inference framework. For this purpose, we need to establish a finite population vector CLT for blocked sample means in randomized block experiments with  multiple treatments, $ \widehat{\bar{  \Y }} = ( \widehat{\bar{Y}}(1),\dots,  \widehat{\bar{Y}}(Q))$.


\section{Joint asymptotic normality}

The finite population vector CLT plays a crucial role in studying the asymptotic properties of treatment effect estimators in randomized experiments \citep{Li2016, Liu2019}. In this section, we first establish a general finite population vector CLT for vector potential outcomes in a randomized block experiment with multiple treatments, and then apply it to $ \widehat { \boldsymbol \tau}_{\unadj} $ to infer the average factorial effects $ { \boldsymbol \tau} $.

\subsection{Finite population vector CLT}

Consider a randomized block experiment with $n$ units, $M$ blocks, and $Q$ treatments. Within block $m$, $\nmq$ ($\nmq \geq 1$) of $\nm$ units are randomly selected and receive  treatment $q$ $(q = 1,\dots,Q)$. For unit $i$, let $R_i(q) \in \mathbb{R}^L$ be an $L$-dimensional ($L\geq 1$) vector of potential outcomes under treatment $q$ and let $\boldsymbol R_i = ( R_i(1)^\T, \dots, R_i(Q)^\T )^\T$ be the $L Q$-dimensional vector of all potential outcomes. For simplicity, we assume that both $L$ and $Q$ are fixed. In the following, $R_i(q) $ can be $Y_i(q)$, $\X_i$, or their transformations. Let us denote the vector of  blocked sample means as $\widehat {\bar{\boldsymbol R}} = ( \widehat{ \bar{R}} (1)^\T, \dots, \widehat {\bar{R}}(Q)^\T )^\T$; we then have the following theorem regarding the mean and covariance of $\widehat {\bar{\boldsymbol R}}$.

\begin{theorem}\label{thm::multiple-var}
In a randomized block experiment with $n$ units, $M$ blocks, and $Q$ treatments, the blocked sample mean  $\widehat {\bar{\boldsymbol R}}$ has mean  $ {\bar{\boldsymbol R}}$ and covariance
\[ \cov ( \widehat {\bar{\boldsymbol R}} ) = \summ \pim^2  \Big[  \textnormal{diag}\Big \{  \frac{1}{\nmq} \S^2_{[m]\R(q)}, q = 1,\dots, Q   \Big \}  - \dfrac{1}{\nm}  \S^2_{[m] \boldsymbol R } \Big],  \]
where ``diag" denotes a block-wise diagonal matrix with its arguments on the diagonal.
\end{theorem}

According to the finite population CLT established in \cite{bickel1984asymptotic} and \cite{Liu2019},  each element of $\widehat {\bar{\boldsymbol R}} $ is asymptotically normal. However, it does not imply the joint asymptotic normality of  $\widehat {\bar{\boldsymbol R}} $, which is required to  determine the asymptotic distribution of $ \widehat { \boldsymbol \tau}_{\unadj}  $. For this purpose, we need the following conditions.



\begin{condition}\label{cond core 0}
For $ m = 1,\dots, M, \ q =1, \dots, Q$, there exist constants $\emq^{\infty}$ and $C_1 \in (0, 0.5)$ independent of $n$,  such that $ C_1 <  \emq^{\infty} <1 - C_1$, and as $ n \rightarrow \infty $,
$   \max_{m=1,\dots,M} \max_{q = 1,\dots,Q}  |  \emq - \emq^{\infty} | \rightarrow 0 . $
\end{condition}

\begin{condition}\label{cond second moment}
\blue{(a) There exists a constant  $C_2 > 0$ independent of $n$, such that
$$
\max_{m=1,\dots,M} \max_{q = 1,\dots,Q} \nm^{-1} \sumim  \|   R_i(q) - \bar{R}_{[m]}(q)  \|_{\infty}^{2 } \leq C_2;
$$
(b) As $n \rightarrow \infty$,
$
\max_{m=1,\dots,M} \max_{i \in [m]} \max_{q = 1,\dots,Q}  \|   R_i(q) - \bar{R}_{[m]}(q)  \|_{\infty}^2 / n \rightarrow 0.
$
}
\end{condition}



Condition~\ref{cond core 0} ensures that the block-specific propensity scores ($\emq$) converge uniformly to constants between zero and one. Condition~\ref{cond second moment} (a) assumes that the block-specific second moments of the potential outcomes are uniformly bounded. Condition~\ref{cond second moment} (b) involves the restriction on the order of the maximum squared distance of the potential outcomes from their block-specific means, which is a typical condition for deriving the finite population CLT; see for example, \citet{Hajek::1961, sen1995hajek, Li2016}, and \citet{Liu2019}.
\blue{Note that the sample size $n\rightarrow \infty$ implies that the number of blocks $M \rightarrow \infty$ and/or the block size $\nm \rightarrow \infty$ for some $m$. Moreover,
these conditions allow for the units to change block membership as $n$ grows.}







\begin{theorem}\label{thm::vclt}
In a randomized block  experiment with $n$ units, $M$ blocks, and $Q$ treatments, under Conditions~\ref{cond core 0} \blue{and \ref{cond second moment}}, if $ n \times \cov  ( \widehat {\bar{\boldsymbol R}} ) $ converges to a finite limit, denoted by $ \boldsymbol  \Sigma$,  then
$ \sqrt{n}  (  \widehat {\bar{\boldsymbol R}} -  \bar{\boldsymbol R}  )  \stackrel{d}{\longrightarrow} \mathcal{N}(  \boldsymbol 0,  \boldsymbol  \Sigma ). $
\end{theorem}


Theorem~\ref{thm::vclt} establishes the joint asymptotic normality of $ \widehat {\bar{\boldsymbol R}}$, which is useful for investigating the asymptotic properties of  general causal estimators in  randomized block experiments with multiple treatments. The conclusion holds for the cases of \blue{only} large blocks, many small blocks, and some combination thereof, provided that the total number of units $n$ tends to infinity, the propensity scores are uniformly bounded between zero and one, and  $ n \times \cov (  \widehat {\bar{\boldsymbol R}}) $ converges to a finite limit. \red{Here and in what follows, we say a block is large if its size is much larger than the number of covariates, and a block is small if its size is smaller than or comparable to the number of covariates.} When $M = 1$ (i.e., in completely randomized experiments with $Q$ treatments), the conclusion of this theorem has been obtained by \cite{Li2016} using the vector form of the Wald--Wolfowitz--Hoeffding theorem for a bi-linear rank statistic under random permutation. Theorem~\ref{thm::vclt} generalizes the result to randomized block experiments with limited requirements on the number of blocks and block sizes.  Theorem~\ref{thm::vclt} also extends the results of \cite{bickel1984asymptotic} and \cite{Liu2019}, from the situation of one-dimensional potential outcomes and two treatments to that of $L$-dimensional vector potential outcomes and multiple treatments. This extension is non-trivial owing to the complex dependence structure between the elements of $ \widehat {\bar{\boldsymbol R}}$ and the lack of the vector form of the Wald--Wolfowitz--Hoeffding theorem for a bi-linear rank statistic under blocked permutation. We obtain this theorem by making use of the techniques to prove the Wald--Wolfowitz--Hoeffding theorem for a linear rank statistic \citep{Hajek::1961, sen1995hajek}, that is, constructing an asymptotically equivalent random variable that is the sum of independent random variables, and then, applying the classical Linderberg--Feller CLT. The detailed proof is given in the Supplementary Material.






\subsection{Joint asymptotic normality of \texorpdfstring{$\widehat { \boldsymbol \tau }_{\unadj}$}{tau}}

As $\widehat { \boldsymbol \tau }_{\unadj}$ is a linear contrast of the blocked sample means $\widehat { \bar{ \boldsymbol  Y}}=  ( \widehat{ \bar{Y}} (1), \dots, \widehat {\bar{Y}}(Q) )^\T$, we can obtain the joint asymptotic normality of $\widehat { \boldsymbol \tau }_{\unadj}$ by applying Theorem~\ref{thm::vclt} to  $\widehat { \bar{ \boldsymbol  Y}}$. For this purpose, we assume the following condition which guarantees the convergence of the covariance, $\cov ( \sqrt{n}  \widehat{\boldsymbol \tau}_\unadj   ) $.




\begin{condition}\label{cond variance}
The weighted variances and covariances, $ \sum^M_{m=1} \pi_{[m]} { S^2_{[m]\Y(q)} }/{ \emq }$ and $\sum^M_{m=1} \pi_{[m]} $ $ S_{[m] \Y(q)\Y(q') }$, $1\leq q,q' \leq Q$, tend to finite limits.
\end{condition}


\begin{theorem}\label{thm::unadj}
If Conditions \ref{cond core 0}, \blue{\ref{cond second moment} for $R_i(q) = Y_i(q)$, and \ref{cond variance}} hold, then $ \widetilde{ \boldsymbol  V}_n (\Y) $ and $\boldsymbol  V_n (\Y) $ have finite limits, denoted by $ \widetilde{\boldsymbol V}( \Y )$ and ${\boldsymbol V}( \Y )$ respectively, and $ \sqrt{n} \{  \widehat{\boldsymbol \tau}_\unadj - \boldsymbol \tau \}   \stackrel{d}{\longrightarrow} \mathcal{N}(  \boldsymbol 0, \widetilde{\boldsymbol V}( \Y ) ). $
Furthermore, if  $  \nmq \geq  2 $ for $m= 1,\dots,M$, $q=1,\dots,Q$,
then, the covariance estimator $  \widehat{ \boldsymbol V}_n(\Y) $ converges in probability to ${\boldsymbol V}( \Y )$ and
$ {\boldsymbol V}( \Y ) -   \widetilde{\boldsymbol V}( \Y )  = \lim_{n \rightarrow \infty} \sum^M_{m=1} \pi_{[m]}\boldsymbol S^2_{[m] \boldsymbol  \tau} \geq 0  . $
\end{theorem}

Theorem~\ref{thm::unadj} establishes the joint asymptotic normality of $  \widehat{\boldsymbol \tau}_\unadj $ and provides an asymptotically conservative estimator for the asymptotic covariance. The covariance estimator is consistent if  $ \boldsymbol S^2_{[m] \boldsymbol  \tau} = 0$, that is, if the unit-level factorial effects are block-specifically additive,
$ \boldsymbol \tau_i =  \boldsymbol c_m $ for all $ i \in [m] $, where $ \boldsymbol  c_m $ is a constant vector. In such a case, Theorem~\ref{thm::unadj} can be used for randomization-based inferences of factorial effects under the additive causal effects assumption, such as conducting  tests or calculating $p$-values under Fisher's sharp null hypothesis. In general cases, Theorem~\ref{thm::unadj}  is useful for constructing large-sample conservative confidence intervals for each factorial effect or confidence regions for joint factorial effects. More specifically, if the limit of ${\boldsymbol V}( \Y )$ is nonsingular, then the probability that $  \widehat{ \boldsymbol V}_n(\Y) $ is nonsingular converges to one. For $\alpha  \in (0,1)$, let $\chi^2_{F_1} ( 1 - \alpha )$ be the $(1 - \alpha)$th quantile of a $\chi^2$ distribution with degrees of freedom $F_1$. \blue{We can then construct a Wald-type confidence region for  $\boldsymbol C \boldsymbol \tau$:}
$$
\Big\{  \boldsymbol  \mu : \  n  ( \boldsymbol C \widehat{\boldsymbol \tau}_\unadj - \boldsymbol  \mu  )^\T  \big\{ \boldsymbol C \widehat{  \boldsymbol V}_n (\Y) \boldsymbol C^\T \big\}^{-1}   ( \boldsymbol C \widehat{  \boldsymbol \tau}_\unadj - \boldsymbol  \mu  )  \leq   \chi^2_{F_1} ( 1 - \alpha )   \Big\},
$$
where $\boldsymbol C \in \mathbb{R}^{F_1 \times F}$ ($F_1 \leq F$) is a constant matrix and has full row rank and the asymptotic coverage rate is at least as large as $1 - \alpha$.  The asymptotic coverage rate equals $1 - \alpha$ if and only if $ \lim_{n \rightarrow \infty} \sum^M_{m=1} \pi_{[m]}\boldsymbol S^2_{[m] \boldsymbol  \tau} = \boldsymbol{0}$.


\begin{remark}
The authors in \cite{aronow2014} proposed a consistent estimator for the sharp bound on the asymptotic variance of the difference-in-means estimator in completely randomized experiments with scalar outcomes and two treatments. Their proposed estimator is generally less conservative than the Neyman-type variance estimator. It will be interesting to extend their results to randomized block factorial experiments.
\end{remark}


\section{Covariate adjustment}

It is widely recognized that adjusting
for the imbalances of baseline covariates can improve the treatment effect estimation efficiency in randomized experiments, including completely randomized experiments \citep{lin2013,Lei2020}, randomized block experiments  \citep{Liu2019}, and completely randomized $2^K$ factorial experiments \citep{lu2016covariate, lu2016randomization}. In this section, we propose four covariate adjustment methods in randomized block $2^K$ factorial experiments according to the number of blocks, block sizes, and propensity scores, and compare their efficiencies with that of the unadjusted estimator.


\subsection{\red{Existence of small blocks, equal propensity scores}}

Covariate adjustment is a standard statistical approach in the analysis of randomized experiments to improve estimation efficiency, following a similar spirit in the survey sampling literature \citep[e.g.,][]{Cassel1976, Sarndal2003}. To increase the estimation accuracy of  the  mean $\bar{Y} (q)$, $q = 1,\dots,Q$,  an adjusted estimator of the form $ \widehat{ \bar{Y}} (q) -  \{ \widehat{\bar{\X}}(q) -  \bar{\X}   \} ^\T \widehat{\boldsymbol \beta}(q)$ is often used to replace the simple blocked sample mean $ \widehat{ \bar{Y}} (q) $, where $\widehat{\boldsymbol \beta}(q)$ is an (estimated) adjusted vector. In completely randomized experiments with two or more treatments, $\widehat{\boldsymbol \beta}(q)$ can be obtained by regressing the observed outcomes on the covariates using the sample under treatment arm $q$. More importantly, under mild conditions, the efficiency gain of estimating the individual  mean $\bar{Y} (q)$ can yield efficiency gain of estimating the average treatment effect, regardless of the correlation structure of $ \widehat{ \bar{Y}} (q) -  \{ \widehat{\bar{\X}}(q) -  \bar{\X}   \} ^\T \widehat{\boldsymbol \beta}(q)$ between treatments \citep{lin2013, lu2016covariate, lu2016randomization}. However, such an efficiency gain is not guaranteed in randomized block  experiments, as shown in the following Theorem~\ref{thm::adj}. In this section, we first study how to obtain the optimal adjusted vector for estimating $\bar{Y} (q)$ with multiple treatments, and then discuss the conditions under which the resulting covariate-adjusted factorial effects estimator is more efficient than the unadjusted estimator.



The optimal adjusted vector $ {\boldsymbol \beta}(q) $ can be obtained by minimizing the variance of $ \widehat{ \bar{Y}} (q) - \{ \widehat{\bar{\X}}(q) -  \bar{\X}  \} ^\T \boldsymbol \beta  $:
\begin{eqnarray}\label{eqn::opt-beta}
 {\boldsymbol \beta}(q)  =  \argmin_{ \boldsymbol \beta }  \var \big[  \widehat{ \bar{Y}} (q) - \{ \widehat{\bar{\X}}(q) -  \bar{\X}  \} ^\T \boldsymbol \beta   \big]
  =  \argmin_{ \boldsymbol \beta }  \frac{1}{n} \summ \pim \frac{1 - \emq }{ \emq } S^2_{[m]\{\Y(q) - \X ^\T  \boldsymbol \beta  \}},
\end{eqnarray}
where the second equality is obtained by applying Theorem~\ref{thm::multiple-var} to the transformed outcomes $Y_i(q) - \X_i ^\T  \boldsymbol \beta$. The optimal adjusted vector $ {\boldsymbol \beta}(q) $ can be consistently estimated by replacing the block-specific  variance in \eqref{eqn::opt-beta} by the corresponding sample variance:
\begin{eqnarray*}
\widehat {\boldsymbol \beta}(q)  =   \argmin_{ \boldsymbol \beta }  \summ  \frac{1 - \emq }{ \emq }  \frac{\pim }{\nmq - 1} \sumim I(Z_i = q) \Big[  Y_i(q) - \hbymq - \{ \boldsymbol X_i - \hbxmq \} ^\t \boldsymbol \beta  \Big]^2.
\end{eqnarray*}
\blue{This is equivalent to performing the following linear regression with weights $\omega_i = \{(1 - \emq) \nm \} / $ $ \{ \emq (\nmq - 1)\} $ for $i \in [m]$ and $Z_i = q$:
\begin{eqnarray}
 Y_i^\obs  =    \sumq \sum_{m=1}^{M} \alpha_{[m]}(q) I(Z_i = q, i \in [m])    + \sum^Q_{q = 1} I(Z_i = q)  (\boldsymbol X_i-\bar{\X})^\T \bbeta(q) + \epsilon_i , \nonumber
\end{eqnarray}
where $Y_i^\obs = \sumq I(Z_i = q) Y_i(q)$ is the observed outcome. Then, $ \widehat {\boldsymbol \beta}(q) $ is equal to the weighted least squares (WLS) estimator of $\bbeta(q)$. Let $\hat \alpha_{[m]}(q)$ be the WLS estimator of $\alpha_{[m]}(q)$.
Replacing $ \widehat{ \bar{Y}} (q) $ in $  \widehat { \boldsymbol \tau }_{\unadj}  $ by $ \summ \pim \hat \alpha_{[m]}(q) =   \widehat{ \bar{Y}} (q) - \{ \widehat{\bar{\X}}(q) -  \bar{\X}  \} ^\T \widehat {\boldsymbol \beta}(q) $,} we obtain a covariate-adjusted average factorial effects estimator of $\boldsymbol \tau$,
\begin{equation*}
 \widehat { \boldsymbol \tau }_{\adj} =  \dfrac{1}{2^{K-1}}   \sum^Q_{q = 1} \dq \Big[ \widehat{ \bar{Y}} (q) - \big \{ \widehat{\bar{\X}}(q) -  \bar{\X} \big \} ^\T \widehat {\boldsymbol \beta}(q) \Big].
\end{equation*}

\begin{remark}
For the case of two treatments $(Q=2)$, \cite{Liu2019} proposed to use the following adjusted vector, $q = 1,2$,
\begin{eqnarray}
\widetilde {\boldsymbol \beta}(q)  =   \argmin_{ \boldsymbol \beta }  \summ  \frac{\pim }{\nmq - 1} \sumim I(Z_i = q) \Big[  Y_i(q) - \hbymq - \{ \boldsymbol X_i - \hbxmq \} ^\t \boldsymbol \beta  \Big]^2. \nonumber
\end{eqnarray}
Note that, when the propensity scores $ \emq $ are equal across blocks, $\widehat {\boldsymbol \beta}(q)$ has the same asymptotic limit as $\widetilde {\boldsymbol \beta}(q) $. Thus, $\widehat {\boldsymbol \beta}(q) $ can be considered extension of $\widetilde {\boldsymbol \beta}(q) $ to general propensity scores and multiple treatments.
\end{remark}

To study the asymptotic property of $ \widehat { \boldsymbol \tau }_{\adj}$, we need to decompose the potential outcomes as follows: $Y_i(q) = \bar Y_{[m]}(q) + (\boldsymbol X_i - \bar{\boldsymbol X}_{[m]})^\t \boldsymbol  \beta (q) + \e_i(q)$, $i \in [m]$, $m= 1,\dots,M$,
where $\e_i(q)$ are (fixed) decomposition errors. It is easy to see that the block-specific mean of $ \e_i(q)$ is zero, i.e., $\bar{\e}_{[m]}(q) = 0$ for each block $m$. In addition, we need the following condition:


\begin{condition}\label{cond covariance x}
The following weighted variances and covariances tend to finite limits:
$$ \summ \frac{ \pim }{ \emq }  \S_{[m]\X\X},\quad \summ \pim \S_{[m]\X\X}, \quad \sum^M_{m=1}  \frac{ \pim }{ \emq }  \S_{[m] \X \Y(q) }, \quad \sum^M_{m=1} \pi_{[m]} \S_{[m] \X \Y(q) }, \quad q=1,\dots,Q,$$
and the limits of the first two matrices and their difference are positive definite.
\end{condition}


Condition~\ref{cond covariance x} ensures that the estimated adjusted vector $\widehat {\boldsymbol \beta}(q) $ converges in probability to the limit of the optimal adjusted vector $ {\boldsymbol \beta}(q) $.
Let $\X  \bbeta = ( \X \bbeta(1), \dots, \X \bbeta(Q) ) $ and $\be_i = (\e_i(1),\dots,\e_i(Q))^\T$. \red{Define $ \widetilde{ \boldsymbol  V}(\be) $ and $  \boldsymbol  {V} (\be) $ similarly to  $ \widetilde{ \boldsymbol  V} (\Y) $ and $\boldsymbol  V (\Y)$ except that $\Y_i$ is replaced by $\be_i$.}



\begin{theorem}\label{thm::adj}
Under Conditions~\ref{cond core 0}, \blue{\ref{cond second moment} for $R_i(q)=Y_i(q), \X_i$, \ref{cond variance}, and \ref{cond covariance x}}, if $  \nmq \geq  2 $ for $m= 1,\dots,M$, $q=1,\dots,Q$,  then $ \widetilde{ \boldsymbol  V}_n (\be) $ and $\boldsymbol  V_n (\be) $ have finite limits, denoted by $ \widetilde{\boldsymbol V}( \be )$ and ${\boldsymbol V}( \be )$ respectively, and $ \sqrt{n} \{  \widehat{\boldsymbol \tau}_\adj - \boldsymbol\tau \}   \stackrel{d}{\longrightarrow} \mathcal{N}(  \boldsymbol 0, \widetilde{ \boldsymbol  V}( \be ) ).$ Furthermore, the difference between the asymptotic covariances of $  \widehat{\boldsymbol \tau}_\unadj$ and $  \widehat{\boldsymbol \tau}_\adj $ is the limit of
\begin{eqnarray}
\label{eqn::adj-unadj}
&&  \widetilde{ \boldsymbol  V}_n ( \X \boldsymbol \beta ) +  2^{-2(K-1)} \summ \pim \sumq \frac{2 }{ \emq }  S_{[m] \{ \X  \boldsymbol  \beta(q) \} \{ \be(q) \} }  \dq  \dq^\T \\
&  & -  2^{-2(K-1)} \summ \pim  \sum_{1\leqslant q, q' \leqslant Q}   \Big\{ S_{[m] \{ \X  \boldsymbol  \beta(q' ) \} \{ \be(q) \} } + S_{[m] \{ \X  \boldsymbol  \beta(q) \} \{ \be(q') \} }  \Big\}   \dq  \dqq^\T.  \nonumber
\end{eqnarray}
\end{theorem}


The first term in the difference between asymptotic covariances \eqref{eqn::adj-unadj} is positive definite, whereas in general, the second and third terms can be either positive definite or negative definite. Thus, the difference between the asymptotic covariances of $  \widehat{\boldsymbol \tau}_\adj $ and $  \widehat{\boldsymbol \tau}_\unadj$ can be either positive or negative definite. Therefore, minimizing the variance of $ \widehat{ \bar{Y}} (q) - \{ \widehat{\bar{\X}}(q) -  \bar{\X}  \}^\T \boldsymbol \beta  $ separately cannot guarantee the reduction of variance for estimating the average factorial effects $\boldsymbol \tau$. This is a significant difference between the performances of covariate adjustment in completely randomized and randomized block experiments. The last two terms are equal to zero in some special cases discussed below:

\begin{itemize}
\item Without blocking ($M = 1$). According to the definition of ${\boldsymbol \beta}(q)$ in \eqref{eqn::opt-beta}, we have $ {\boldsymbol \beta}(q)   =   \argmin_{ \boldsymbol \beta } \{  { ( 1 - e_{[1]q} ) } / { e_{[1]q} } \} S^2_{[1]\{\Y(q) - \X ^\T  \boldsymbol \beta  \}}.$
Then, the decomposition errors $\e_i(q)$ are orthogonal to the covariates $\X_i$ in the sense that $\S_{[1] \X \be(q) } =   \boldsymbol 0$, which implies
$ S_{[1] \{ \X  \boldsymbol  \beta(q) \} \{ \be(q') \} }  = \S_{[1] \{ \X   \} \{ \be(q') \} } ^\T \boldsymbol  \beta(q) = 0,$ $  1 \leq q, q' \leq Q.$
Thus, the last two terms in \eqref{eqn::adj-unadj} are zero. In other words, when blocking is not used in the design stage, we can adjust for covariate imbalances for potential outcomes under each treatment arm separately by minimizing the variance of the adjusted estimator, which guarantees the efficiency gain of the resulting covariate-adjusted factorial effect estimator $  \widehat{\boldsymbol \tau}_\adj $.

\item Equal propensity scores. When $\emq = e_q$ for $m=1,\dots,M$, then according to the  definition of ${\boldsymbol \beta}(q)$  in \eqref{eqn::opt-beta}, we have $ {\boldsymbol \beta}(q)   =   \argmin_{ \boldsymbol \beta }  \summ \pim S^2_{[m]\{\Y(q) - \X ^\T  \boldsymbol \beta  \}}.$
Thus, the decomposition errors $\e_i(q)$ are orthogonal to the covariates $\X_i$ in the sense that $\summ \pim \S_{[m] \X \be(q) } =  \boldsymbol 0$, which again implies that the last two terms in \eqref{eqn::adj-unadj} are equal to zero. Therefore,  the covariate-adjusted factorial effect estimator $\widehat{\boldsymbol \tau}_\adj$ is asymptotically more efficient than, or at least as efficient as, the unadjusted estimator $\widehat{\boldsymbol \tau}_\unadj$. As equal propensity scores are common  in practice, we discuss this special case in more detail.
\end{itemize}

We define residuals as follows:
$ \widehat \e_i (q)  = Y_i(q) -  \hbymq -  \{ \X_i -  \hbxmq \} ^\T \widehat {\boldsymbol \beta}(q) . $
Similar to the arguments for $  \widehat{\boldsymbol \tau}_\unadj$, the asymptotic covariance of $  \widehat{\boldsymbol \tau}_\adj$ can be estimated by  $ \widehat { \boldsymbol V}_n ( \widehat \be )$, \red{which is defined similarly to $\widehat { \boldsymbol V}_n ( \Y )$ except that $\Y_i$ is replaced by $\widehat \be_i = ( \widehat \e_i (1),\dots, \widehat \e_i (Q))^\T$.}


\begin{corollary}\label{coro adj var}
Under Conditions~\ref{cond core 0}, \blue{\ref{cond second moment} for $R_i(q)=Y_i(q), \X_i$, \ref{cond variance}, and \ref{cond covariance x}}, if $  \nmq \geq  2 $  for $m= 1,\dots,M$, $q=1,\dots,Q$, then
$  \widehat { \boldsymbol V}_n ( \widehat \be ) \stackrel{p}{\longrightarrow} {\boldsymbol V}( \be )  \geq   \widetilde{\boldsymbol V}( \be ) .$
Furthermore, if $\emq^{\infty} = e_q$ ($m=1,\dots,M$), then the difference between the asymptotic covariances of $  \widehat{\boldsymbol \tau}_\unadj$ and $  \widehat{\boldsymbol \tau}_\adj $ is the limit of $ \widetilde{ \boldsymbol  V}_n ( \X \boldsymbol \beta )  \geq 0, $
and the difference between the limits of the covariance estimators $  \widehat { \boldsymbol V}_n ( \boldsymbol Y  ) $ and $  \widehat { \boldsymbol V}_n ( \widehat \be ) $ is the limit of $ { \boldsymbol  V}_n ( \X \boldsymbol \beta )  \geq \boldsymbol{0}. $
\end{corollary}

\begin{remark}
\blue{$  \widehat{\boldsymbol \tau}_\adj $ can be viewed as $\widehat{\boldsymbol \tau}_\unadj$ adjusting for  $\{\hbxq, q = 1,\dots,Q \}$.
Let $\widehat{\boldsymbol \tau}_{\X,f}  = {2^{-(K-1)}}   \sum^Q_{q = 1} g_{f,q}  \widehat {\bar{\X}}(q)$, $f=1,\dots,F$, be the observed factorial effects of the covariates.
If $\emq^{\infty} = e_q$, it can be shown that $\sumq e_q \hbxq = \boldsymbol  0 $ and  $\{\hbxq, q = 1,\dots,Q \}$ is a linear transformation of $ \widehat{\btau}_{\X} =(\widehat{\boldsymbol \tau}_{\X,1}^\T,\dots,\widehat{\boldsymbol \tau}_{ \X, F}^\T)^\T$ and vice versa. Moreover,
$$
\lim_{n \rightarrow \infty} \cov( \widehat{\boldsymbol \tau}_\adj )  = \lim_{n \rightarrow \infty} \min_{\bGamma } E ( \widehat{\boldsymbol \tau}_\unadj - \boldsymbol \tau -  \bGamma^\t \widehat{\btau}_{\X} )  ( \widehat{\boldsymbol \tau}_\unadj - \boldsymbol \tau -  \bGamma^\t \widehat{\btau}_{\X} )^\T.
$$
That is,  $ \widehat{\boldsymbol \tau}_\adj $ is equivalent to projecting $\widehat{\boldsymbol \tau}_\unadj$ on $ \widehat{\btau}_{\X}$ for the case of equal propensity scores across blocks.}
\end{remark}

According to Corollary~\ref{coro adj var}, the covariance estimator $ \widehat { \boldsymbol V}_n ( \widehat \be )$ is generally conservative, and it is consistent if and only if the limit of the weighted average of the block-specific covariances of the unit-level factorial effects is  zero, i.e., $ \lim_{n \rightarrow \infty} \sum^M_{m=1} \pi_{[m]}\boldsymbol S^2_{[m] \boldsymbol  \tau} = 0 $. As with $  \widehat{\boldsymbol \tau}_\unadj$, \blue{we can construct a Wald-type confidence region for $\boldsymbol C \boldsymbol \tau$:}
$$
\blue{\Big \{  \boldsymbol  \mu : \  n  ( \boldsymbol C \widehat{\boldsymbol  \tau}_\adj - \boldsymbol  \mu  )^\T  \big\{ \boldsymbol C  \widehat{ \boldsymbol V}_n (  \widehat \be  ) \boldsymbol C^\T \big \} ^{-1}  ( \boldsymbol C \widehat{\boldsymbol \tau}_\adj - \boldsymbol  \mu  )  \leq   \chi^2_{F_1} ( 1 - \alpha )  \Big \},}
$$
whose asymptotic coverage rate is at least as large as  $1 - \alpha$. Furthermore, when the propensity scores are asymptotically the same across blocks ($\emq^{\infty} = e_q$), both the asymptotic covariance and covariance estimator of the covariate-adjusted average factorial effect estimator $  \widehat{\boldsymbol \tau}_\adj $ are less than, or equal to those of the unadjusted estimator $ \widehat{\boldsymbol \tau}_\unadj$. Thus, it is generally more efficient to conduct inferences for $\boldsymbol \tau$ based on $ \widehat{\boldsymbol \tau}_\adj$ and $  \widehat { \boldsymbol V}_n ( \widehat \be )$.

\subsection{\red{Existence of small blocks, unequal propensity scores}}

It is not always possible to ensure the same propensity scores across blocks because of practical restrictions. For example, consider an experiment with 10 men and 11 women; if blocked by gender, it is impossible to make the propensity scores equal across blocks. To improve the estimation efficiency of $\boldsymbol  \tau$ for unequal propensity scores, we propose \blue{two covariate adjustment methods}, from a conditional inference perspective. Conditional inference is an influential idea in statistics that began with the original ideas of Fisher \citep{Fisher1959}.



\subsubsection{Conditional on a single factorial effect of the covariates}

\blue{In this section, we introduce the idea of conditional inference for estimating each factorial effect $ \tau_f = {2^{-(K-1)}}  \sum^Q_{q = 1} g_{f,q}  \bar{Y}(q)$, $f=1,\dots,F$,  $\sumq g_{f,q} = 0$.} Generally, we can improve the inference efficiency of $\tau_f$ conditional on \blue{$\widehat{\boldsymbol \tau}_{\X,f}$. Applying Theorem~\ref{thm::vclt} to $(\Y, \X)$ yields $\sqrt{n} (\widehat{ \tau}_{f,\unadj} -  \tau_f, (\widehat{\boldsymbol \tau}_{\X,f} - {\boldsymbol \tau}_{\X,f})^\T )^\T  \stackrel{d}{\longrightarrow}  \mathcal{N} ( \boldsymbol 0,  \boldsymbol \Sigma_{f}  )  $, where $ {\boldsymbol \tau}_{\X,f} =  {2^{-(K-1)}}   \sum^Q_{q = 1} g_{f,q}  {\bar{\X}}(q) = \boldsymbol 0$ and}
\begin{eqnarray}
\blue{ \boldsymbol \Sigma_{f} = \lim_{n \rightarrow \infty}
\left(
\begin{array}{cc}
 \Sigma_{f, n, \boldsymbol \tau  \boldsymbol \tau} &  \boldsymbol \Sigma_{f, n, \boldsymbol \tau \X} \\
 \boldsymbol \Sigma_{f, n,\X \boldsymbol \tau} &  \boldsymbol \Sigma_{f, n, \X \X}
\end{array}
\right) = \lim_{n \rightarrow \infty}
\left(
\begin{array}{cc}
 \var ( \sqrt{n}   \widehat{ \tau}_{f,\unadj}  ) &  \cov (  \sqrt{n}   \widehat{ \tau}_{f,\unadj}  ,  \sqrt{n}  \widehat{\boldsymbol \tau}_{\X,f}   ) \\
 \cov (  \sqrt{n}    \widehat{\boldsymbol \tau}_{\X,f},  \sqrt{n} \widehat{ \tau}_{f,\unadj}   ) &   \cov ( \sqrt{n}  \widehat{\boldsymbol \tau}_{\X,f}  )
\end{array}
\right).} \nonumber
\end{eqnarray}
Let \blue{$ \boldsymbol \gamma_f = \boldsymbol \Sigma_{f, n, \X \X}^{-1}  \boldsymbol \Sigma_{f, n,  \X \boldsymbol \tau}$.} Then,  conditional on $ \sqrt{n} (  \widehat{\boldsymbol \tau}_{\X, f} - {\boldsymbol \tau}_{\X, f} ) $,  $\sqrt{n} \widehat{ \tau}_{f,\unadj}  $ is asymptotically normal with mean
$ \sqrt{n} \boldsymbol \gamma_f^\T ( \widehat{\boldsymbol \tau}_{\X, f}  - {\boldsymbol \tau}_{\X, f} )  $ and variance \blue{
$\Sigma_{f, n, \boldsymbol \tau \boldsymbol \tau} - \boldsymbol \Sigma_{f,n, \boldsymbol \tau \X} \boldsymbol \Sigma_{f, n, \X \X}^{-1} $ $ \boldsymbol \Sigma_{f, n, \X \boldsymbol \tau}  \leq \Sigma_{f, n, \boldsymbol \tau \boldsymbol \tau}.$}
Therefore, removing the bias, \blue{$ \boldsymbol \gamma_f ^\T   ( \widehat{\boldsymbol \tau}_{\X, f} - {\boldsymbol \tau}_{\X, f} ) $}, will result in a consistent and more accurate estimator,
\begin{equation*}
\blue{\widehat \tau_{f,\cond}   =   \widehat{ \tau}_{f,\unadj}  -  \widehat{ \boldsymbol \gamma}_f^\T ( \widehat{\boldsymbol \tau}_{\X, f} - {\boldsymbol \tau}_{\X, f} ) =  \sumq g_{f,q} \Big[ \hbyq  - \big \{ \hbxq - \bx \big \} ^\T \widehat {\boldsymbol \gamma}_f \Big],}
\end{equation*}
where \blue{$\widehat {\boldsymbol \gamma}_f$} is a consistent estimator of the adjusted vector \blue{$  \boldsymbol \gamma_f$}. The adjusted average factorial effects estimator is \blue{$\widehat { \boldsymbol \tau }_{\cond} = (\widehat \tau_{1,\cond}, \dots, \widehat \tau_{F,\cond} )^\T$. Let $\s_{[m]\X\X(q)}$ be the sample covariance of $\X$ under treatment combination $q$ in block $m$.}

\begin{prop}\label{prop gamma}
Under Conditions~\ref{cond core 0}, \blue{\ref{cond second moment} for $R_i(q)=Y_i(q), \X_i$, \ref{cond variance}, and \ref{cond covariance x}},  if $  \nmq \geq  2 $  for $m= 1,\dots,M$, $q=1,\dots,Q$, then
\[ \blue{ \boldsymbol \Sigma_{f, n, \X \boldsymbol \tau} = 2^{-2(K-1)}  \summ \pim \sumq \frac{1}{\emq} \S_{[m]\X \Y(q)}   , \quad \boldsymbol \Sigma_{f, n, \X \X} = 2^{-2(K-1)}    \summ \pim \sumq \frac{1}{\emq} \S_{[m]\X\X}.} \]
Furthermore, a consistent estimator of \blue{$\boldsymbol \gamma_f $} is
\[ \blue{\widehat {\boldsymbol \gamma}_f} =  \Big( \summ \pim \sumq \frac{1}{\emq} \s_{[m]\X\X(q)}   \Big)^{-1} \Big(   \summ \pim \sumq \frac{1}{\emq} \s_{[m]\X \Y(q)}   \Big).  \]
\end{prop}


\blue{Proposition~\ref{prop gamma} implies that $\boldsymbol \gamma_f = \boldsymbol \gamma$ does not depend on $f$. Hence, we can use $\widehat {\boldsymbol \gamma} = \widehat {\boldsymbol \gamma}_f $ to improve the estimation efficiency of all of the elements of factorial effects $\boldsymbol \tau$. Moreover, $\widehat {\boldsymbol \gamma}$ is equal to the WLS estimator of the coefficients of $\X_i$ in the following linear regression with weights $\omega_{i,\cond} =  \nm  / $ $ \{ \emq (\nmq - 1)\} $ for $i \in [m]$ and $Z_i = q$:
\begin{eqnarray}
 Y_i^\obs  =    \sumq \sum_{m=1}^{M} \alpha_{[m]}(q) I(Z_i = q, i \in [m])    +  (\boldsymbol X_i-\bar{\X})^\T \bgamma + \epsilon_{i,\cond}. \nonumber
\end{eqnarray}
Note that the weights are different from those used in $\widehat { \boldsymbol \tau }_{\adj}$. Let $\hat \alpha_{[m],\cond}(q)$ be the WLS estimator of $\alpha_{[m]}(q) $. Then, $\widehat { \boldsymbol \tau }_{\cond}$ is equal to $  \widehat { \boldsymbol \tau }_{\unadj}  $ with
 $ \widehat{ \bar{Y}} (q) $ being replaced by $ \summ \pim \hat \alpha_{[m],\cond}(q) =   \widehat{ \bar{Y}} (q) - \{ \widehat{\bar{\X}}(q) -  \bar{\X}  \} ^\T \widehat {\boldsymbol \gamma} $.
}


\begin{remark}\label{rem::projection cond}
By definition and simple algebra, the adjusted vector $  \bgamma $ can be interpreted as a projection coefficient that  minimizes the variance of $   \sum^Q_{q = 1} g_{f,q} [ \widehat{ \bar{Y}} (q) - \big \{ \widehat{\bar{\X}}(q) -  \bar{\X} \big \} ^\T {\bgamma^* } ] $ with respect to $\bgamma^* $. That is,
\[  \bgamma =   \argmin_{ \bgamma^*}  \var \Big(  \sum^Q_{q = 1} g_{f,q} \big[ \widehat{ \bar{Y}} (q) - \big \{ \widehat{\bar{\X}}(q) -  \bar{\X} \big \} ^\T {\bgamma^* } \big]  \Big).  \]
Moreover, according to Proposition~\ref{prop gamma}, $ \widehat {\boldsymbol \gamma} \stackrel{p}{\longrightarrow}  \bgamma$. Thus, \blue{$ \widehat { \tau }_{f,\cond} $} is \blue{equivalent to projecting} \blue{$\widehat{ \tau}_{f,\unadj}$ on $ \widehat{\boldsymbol   \tau}_{\X,f}$ and has} the smallest variance among the class of estimators that have the same form.
\end{remark}

To investigate the asymptotic property of $  \widehat { \boldsymbol \tau }_{\cond} $, we define the decomposition errors and  residuals as follows: for $i \in [m]$,
$ \err_{i}(q) =  Y_i(q) - \bar Y_{[m]}(q)  -  (\boldsymbol X_i - \bar{\boldsymbol X}_{[m]}) ^\T  \boldsymbol \gamma   $ and
$ \hat \err_{i}(q) =  Y_i(q) -  \hbymq  -   \big \{ \X_i  - \hbxmq \big \}  ^\T \widehat {  \boldsymbol \gamma }. $ \red{Let $\berr_i=(\err_i(1),...,\err_i(Q))^{\T}$ and $\widehat\berr_i=(\hat\err_i(1),...,\hat\err_i(Q))^{\T}$. Define $\widetilde{ \boldsymbol  V}(\berr)$, $\boldsymbol  {V}(\berr)$, and $\widehat { \boldsymbol V}_n ( \widehat \berr )$ similarly to $\widetilde{ \boldsymbol  V}(\Y)$, $\boldsymbol  V (\Y)$, and $\widehat { \boldsymbol V}_n ( \Y )$ except that $\Y_i$ is replaced by $\berr_i$ and $\widehat \berr_i$, respectively.}




\begin{theorem}\label{thm::adj2}
Under Conditions~\ref{cond core 0}, \blue{\ref{cond second moment} for $R_i(q)=Y_i(q), \X_i$, \ref{cond variance}, and \ref{cond covariance x}}, if $  \nmq \geq  2 $  for $m= 1,\dots,M$, $q=1,\dots,Q$,  then $ \widetilde{ \boldsymbol  V}_n (\berr) $ and $\boldsymbol  V_n (\berr) $ have finite limits, denoted by $ \widetilde{\boldsymbol V}(\berr)$ and ${\boldsymbol V}(\berr)$ respectively,
$ \sqrt{n} \{  \widehat{\boldsymbol \tau}_\cond - \boldsymbol \tau \}   \stackrel{d}{\longrightarrow} \mathcal{N}(  \boldsymbol 0, \widetilde{ \boldsymbol  V}(\berr) ) $, and $ \widehat { \boldsymbol V}_n ( \hat \berr ) \stackrel{p}{\longrightarrow} {\boldsymbol V}(\berr)  \geq   \widetilde{\boldsymbol V}(\berr) $. Furthermore,  the difference between the asymptotic covariances of $  \widehat{\boldsymbol \tau}_\unadj $ and $  \widehat{\boldsymbol \tau}_\cond$ is the limit of
\[  \widetilde{\V}_n( \X \bgamma ) +  2^{-2(K-1)} \summ \pim \sumq \dfrac{2}{\emq} S_{[m] \{ \X\boldsymbol \bgamma \}  \{\berr(q) \} }   \dq \dq^\t , \]
and the asymptotic difference between  the covariance estimators $  \widehat { \boldsymbol V}_n ( \boldsymbol Y  ) $ and $  \widehat { \boldsymbol V}_n ( \hat \berr ) $ is the limit of
\[  {\V}_n( \X \bgamma )  + 2^{-2(K-1)} \summ \pim \sumq \dfrac{2}{\emq}  \Big\{    S_{[m] \{ \X\boldsymbol \bgamma \}  \{\berr(q) \} }  \Big\} \dq \dq^\t .  \]
\end{theorem}


Theorem~\ref{thm::adj2} implies that $ \widehat{\boldsymbol \tau}_\cond$ is consistent and jointly asymptotically normal, and that its asymptotic covariance can be conservatively estimated. Similar to $  \widehat{\boldsymbol \tau}_\unadj$ and $ \widehat{\boldsymbol \tau}_\adj$, an asymptotically conservative Wald-type confidence region for \blue{ $\boldsymbol C \boldsymbol \tau$} is
$$
\blue{\Big\{  \boldsymbol  \mu : \  n  ( \boldsymbol C \widehat{\boldsymbol \tau}_\cond - \boldsymbol  \mu  )^\T  \big\{ \boldsymbol C  \widehat{ \boldsymbol V}_n (   \hat \berr   ) \boldsymbol C^\T \big\}^{-1}   (  \boldsymbol C \widehat{\boldsymbol \tau}_\cond - \boldsymbol  \mu  )  \leq   \chi^2_{F_1} ( 1 - \alpha )   \Big\}.}
$$

To compare the efficiencies of $ \widehat{\boldsymbol \tau}_\cond$ and  $  \widehat{\boldsymbol \tau}_\unadj$,  according to Proposition~\ref{prop gamma}, the adjusted vector  $ \bgamma  =   \boldsymbol \Sigma_{f, n, \X \X}^{-1}  \boldsymbol \Sigma_{f, n, \X \tau} =  \argmin_{ \boldsymbol \beta }   \summ \pim \sumq ({1 }/{ \emq }) S^2_{[m]\{ \Y(q) - \X ^\T  \boldsymbol \beta  \}}.$
Thus, the decomposition errors $\err_i(q)$ are orthogonal to the covariates in the following sense:
$ \summ \pim  \sumq  ({1}/{\emq}) \S_{[m] \X \berr(q) } $ $ =  \boldsymbol 0, $
which implies
$ \summ \pim  \sumq  ({1}/{\emq}) S_{[m] \{ \X \boldsymbol \bgamma \}  \{\berr(q) \} }  = 0. $
\begin{itemize}
\item When there is only one factor, i.e., $K=1$ and $Q=2$, we have $\dq \dq^\t  = 1$ for $q = 1, 2$. Thus,
$$
2^{-2(K-1)} \summ \pim \sumq \dfrac{2}{\emq}  S_{[m] \{ \X\boldsymbol \bgamma \}  \{\berr(q) \} } \dq \dq^\t = 0.
$$
Therefore, both the point estimator $\widehat{\boldsymbol \tau}_\cond$ and covariance estimator $ \widehat { \boldsymbol V}_n ( \hat \berr ) $ are no worse than $\widehat{\boldsymbol \tau}_\unadj$ and $ \widehat { \boldsymbol V}_n ( \Y )$. Furthermore, for equal propensity scores, $ \widehat{\boldsymbol \tau}_\cond$ is asymptotically equivalent to $   \widehat{\boldsymbol \tau}_\adj $ (as implied by Theorems~\ref{thm::adj} and \ref{thm::adj2}). In contrast, for unequal propensity scores, $  \widehat{\boldsymbol \tau}_\adj$ may hurt the precision when compared to the unadjusted estimator,
whereas $  \widehat{\boldsymbol \tau}_\cond$ does not.
\item When $K > 1$,  as the diagonal elements of $\dq \dqq^\T$ are all equal to one, the diagonal elements of the differences between the asymptotic covariances and the limits of the covariance estimators of $  \widehat{\boldsymbol \tau}_\unadj$ and $ \widehat{\boldsymbol \tau}_\cond $ are greater than or equal to zero. Therefore, for each factorial effect $\tau_f$ ($f=1,\dots,F$), the covariate-adjusted estimator $\widehat{ \tau}_{f, \cond}$ generally improves the estimation efficiency, and its variance estimator  $  \{ \widehat { \boldsymbol V}_n ( \hat \berr ) \}_{ff} $ (the $f$th diagonal element of  $  \widehat { \boldsymbol V}_n ( \hat \berr ) $)  is no worse than that of the unadjusted estimator, even when the propensity scores differ across blocks.
\red{However, the difference between the asymptotic covariances of $\widehat{\boldsymbol \tau}_\unadj$ and $\widehat{\boldsymbol \tau}_\cond$ is not always positive semidefinite. Hence, for some $\boldsymbol C$, $\boldsymbol C \widehat{\boldsymbol \tau}_\cond$ may be worse than $\boldsymbol C \widehat{\boldsymbol \tau}_\unadj$.}
\end{itemize}


\subsubsection{Conditional on all factorial effects of the covariates}

\blue{The joint efficiencies of $  \widehat{\boldsymbol \tau}_\cond$ and $   \widehat{\boldsymbol \tau}_\unadj $ are not ordered in an unambiguous manner. To address this issue and further improve the efficiency of $ \widehat{\boldsymbol \tau}_\cond$, we propose another estimator  conditional on all of the observed factorial effects of the covariates: $ \widehat{\btau}_{\X} =(\widehat{\boldsymbol \tau}_{\X,1}^\T,\dots,\widehat{\boldsymbol \tau}_{\X, F}^\T)^\T.$
Applying Theorem~\ref{thm::vclt} to $(\Y, \X)$,
$\sqrt{n} (\widehat{ \boldsymbol \tau}_\unadj -  \boldsymbol \tau, (\widehat{\boldsymbol \tau}_{\X} - {\boldsymbol \tau}_{\X})^\T )^\T  \stackrel{d}{\longrightarrow}  \mathcal{N} ( \boldsymbol 0,  \boldsymbol \Sigma  )  $, where $ {\boldsymbol \tau}_{\X} =   \boldsymbol 0$ and
\begin{eqnarray}
 \boldsymbol \Sigma
= \lim_{n \rightarrow \infty}
\left(
\begin{array}{cc}
 \boldsymbol \Sigma_{ n, \boldsymbol \tau \boldsymbol \tau} &  \boldsymbol \Sigma_{ n, \boldsymbol \tau \X} \\
 \boldsymbol \Sigma_{n,\X \boldsymbol \tau} &   \boldsymbol \Sigma_{ n, \X \X}
\end{array}
\right) = \lim_{n \rightarrow \infty}
\left(
\begin{array}{cc}
 \cov ( \sqrt{n}   \widehat{ \boldsymbol \tau}_\unadj  ) &  \cov (  \sqrt{n}   \widehat{ \boldsymbol \tau}_\unadj  ,  \sqrt{n}  \widehat{\boldsymbol \tau}_{\X}   ) \\
 \cov (  \sqrt{n}    \widehat{\boldsymbol \tau}_{\X},  \sqrt{n} \widehat{  \boldsymbol \tau}_\unadj  ) &   \cov ( \sqrt{n}  \widehat{\boldsymbol \tau}_{\X}  )
\end{array}
\right). \nonumber
\end{eqnarray}
Note that $\boldsymbol \Sigma_{ n, \boldsymbol \tau \boldsymbol \tau} = \widetilde { \boldsymbol V }_n (\Y)$, which can be conservatively estimated by $\widehat{ \boldsymbol \Sigma}_{ n, \boldsymbol \tau \boldsymbol \tau} =  \widehat { \boldsymbol V }_n (\Y)$. Similar to Proposition~\ref{prop gamma}, $\boldsymbol \Sigma_{n,\X \boldsymbol \tau}$ and $\boldsymbol \Sigma_{ n, \X \X}$ can be consistently estimated by
$$
\widehat{ \boldsymbol \Sigma}_{n,\X \boldsymbol \tau} = 2^{-2(K-1)}   \summ \pim \sumq  \dfrac{1}{\emq} \dq\dq^\t \otimes  \s_{[m]\X\Y(q)},
$$
$$
\widehat{ \boldsymbol \Sigma}_{ n, \X \X} = 2^{-2(K-1)}   \summ \pim \sumq \dfrac{1}{\emq} \dq \dq^\t \otimes \s_{[m]\X\X(q)},
$$
where $\otimes$ denotes the Kronecker product of two matrices.
Thus, $ \widehat {\boldsymbol \Gamma} = \widehat{ \boldsymbol \Sigma}_{ n, \X \X}^{-1} \widehat{ \boldsymbol \Sigma}_{n,\X \boldsymbol \tau}  $ is a consistent estimator of  $\blue{ \boldsymbol \Gamma = \boldsymbol \Sigma_{n, \X \X}^{-1}  \boldsymbol \Sigma_{n, \X \boldsymbol \tau}}$. Then, conditional on $   \sqrt{n}\widehat{\btau}_{\X}$,  we obtain a more efficient estimator
$
\blue{\widehat \btau_{\condfull} }  =   \widehat{ \btau}_\unadj  - \widehat{ \boldsymbol \Gamma}^\T   (\widehat{\btau}_{\X} - \btau_{\X}),
$
which is equivalent to projecting $ \widehat{ \btau}_\unadj $ on $ \widehat{\btau}_{\X}$.}

\begin{theorem}\label{thm::adj22}
Under Conditions~\ref{cond core 0}, \ref{cond second moment} for $R_i(q)=Y_i(q), \X_i$, \ref{cond variance}, and \ref{cond covariance x}, if $\lim_{n \rightarrow \infty} \boldsymbol \Sigma_{ n, \X \X} > 0$ and $  \nmq \geq  2 $  for $m= 1,\dots,M$, $q=1,\dots,Q$,  then
$ \sqrt{n} \{  \widehat{\boldsymbol \tau}_\condfull - \boldsymbol \tau \}   \stackrel{d}{\longrightarrow} \mathcal{N}(  \boldsymbol 0, \widetilde{ \boldsymbol  V}_{\condfull} )$, where $\widetilde{ \boldsymbol  V}_{\condfull} = \lim_{n \rightarrow \infty} \{ \widetilde { \boldsymbol V }_n (\Y) - \boldsymbol \Sigma_{n,\X \boldsymbol \tau}^\T \boldsymbol \Sigma_{ n, \X \X}^{-1} \boldsymbol \Sigma_{n,\X \boldsymbol \tau} \} \leq \widetilde{ \boldsymbol V} (\Y)  $, and
$$
\widehat{\boldsymbol  V}_{n,\condfull} = \widehat{ \boldsymbol V }_n(\Y) - \widehat{ \boldsymbol \Sigma}_{n,\X \boldsymbol \tau}^\T   \widehat{ \boldsymbol \Sigma}_{ n, \X \X}^{-1} \widehat{ \boldsymbol \Sigma}_{n,\X \boldsymbol \tau}  \stackrel{p}{\longrightarrow} \boldsymbol V(\Y) - \lim_{n \rightarrow \infty}  \boldsymbol \Sigma_{n, \X \boldsymbol \tau}^\T \boldsymbol \Sigma_{n, \X \X}^{-1}  \boldsymbol \Sigma_{n, \X \boldsymbol \tau} \geq  \widetilde{ \boldsymbol  V}_{\condfull}.
$$
Furthermore, if $\emq = e_{q}$ for $m= 1,\dots,M$ and $q=1,\dots,Q$, then,  $\widehat \btau_{\condfull}$ is asymptotically equivalent to $ \widehat \btau_{\adj}$, i.e., $\widetilde{ \boldsymbol  V}_{\condfull} = \widetilde{\boldsymbol V}( \be )$.
\end{theorem}


\blue{Theorem~\ref{thm::adj22} implies that $ \widehat{\boldsymbol \tau}_\condfull$ is consistent and jointly asymptotically normal, and that its asymptotic covariance can be conservatively estimated. Moreover, both the asymptotic covariance and covariance estimator of $ \widehat{\boldsymbol \tau}_\condfull$  are less than or equal to those of $\widehat{\boldsymbol \tau}_\unadj$, regardless of whether the propensity scores are the same across blocks.
According to Theorem~\ref{thm::adj22}, an asymptotically conservative  confidence region for $\boldsymbol C \boldsymbol \tau$ is
$$
\Big\{  \boldsymbol  \mu : \  n  ( \boldsymbol C \widehat{\boldsymbol \tau}_\condfull - \boldsymbol  \mu  )^\T   \big( \boldsymbol C  \widehat{\boldsymbol  V}_{n,\condfull}  \boldsymbol C^\T  \big)^{-1}  ( \boldsymbol C \widehat{\boldsymbol \tau}_\condfull - \boldsymbol  \mu  )  \leq   \chi^2_{F_1} ( 1 - \alpha )   \Big\},
$$
whose area is asymptotically smaller than or equal to that of the confidence region constructed by $\widehat{\boldsymbol \tau}_\unadj$ and $\widehat{  \boldsymbol V}_n (\Y) $.
}

\subsection{\blue{Only} large blocks}
The above three covariate adjustment methods  use the block-common adjusted vectors, that is, they pool together the units of all blocks under each treatment arm.  As the factorial effects $\boldsymbol \tau = \sum^M_{m=1} \pi_{[m]} \boldsymbol \tau_{[m]} $ are the weighted average of block-specific factorial effects, and because randomization is conducted independently across blocks,  it may be more efficient to perform block-specific covariate adjustment when there are  \blue{only} large blocks.  More precisely, we can define  block-specific optimal adjusted vectors as follows: for $q=1,\dots,Q$,
\begin{eqnarray*}\label{eqn::opt-beta-m}
 {\boldsymbol \beta}_{[m]}(q) & = & \argmin_{ \boldsymbol \beta }  \var \big[  \widehat{ \bar{Y}}_{[m]} (q) - \{ \widehat{\bar{\X}}_{[m]} (q) -  \bar{\X}_{[m]}  \} ^\T \boldsymbol \beta   \big]  = \S_{[m]\X \X}^{-1} \S_{[m]\X \Y(q)}. \nonumber
\end{eqnarray*}
The optimal adjusted vector $ {\boldsymbol \beta}_{[m]}(q) $ can be consistently estimated by the corresponding sample quantity (regression of $Y_i^\obs$ on $\X_i$ under each treatment arm and each block),
\begin{eqnarray}
\widehat {\boldsymbol \beta}_{[m]}(q)  =   \s_{[m]\X \X (q)}^{-1} \s_{[m]\X \Y(q)}. \nonumber
\end{eqnarray}
Then, the block-specific covariate-adjusted factorial effect estimator can be defined as
\begin{equation*}
 \widehat { \boldsymbol \tau }_{\adjtwo} =  \dfrac{1}{2^{K-1}}   \sum^Q_{q = 1} \dq   \summ \pim  \Big[  \hbymq - \big \{ \hbxmq -  \bar{\X}_{[m]} \big \} ^\T \widehat {\boldsymbol \beta}_{[m]}(q) \Big].
\end{equation*}
\blue{Equivalently, $\widehat { \boldsymbol \tau }_{\adjtwo}$ can be obtained  using the following linear regression:
\begin{eqnarray}
 Y_i^\obs  =    \sumq \sum_{m=1}^{M} \alpha_{[m]}(q) I(Z_i = q, i \in [m])    + \sumq \sum_{m=1}^{M} I(Z_i = q, i \in [m]) (\boldsymbol X_i-\boldsymbol{\bar X}_{[m]})^\T \bbeta_{[m]}(q) + \epsilon_{i,\adjtwo}. \nonumber
\end{eqnarray}
Then, $\widehat {\boldsymbol \beta}_{[m]}(q)$ is the ordinary least squares (OLS) estimator of $\bbeta_{[m]}(q)$ and $\widehat { \boldsymbol \tau }_{\adjtwo}$ is equal to $  \widehat { \boldsymbol \tau }_{\unadj}  $ with
 $ \widehat{ \bar{Y}}_{[m]} (q) $ being replaced by the OLS estimator of $\alpha_{[m]}(q)$. Moreover, $\widehat { \boldsymbol \tau }_{\adjtwo} $ is equivalent to projecting  $  \widehat { \boldsymbol \tau }_{\unadj}  $ on
$ \{ \hbxmq, q = 1,\dots,Q,m=1,\dots,M \} $. Because both $ \{ \hbxq, q = 1,\dots,Q \} $ and $\widehat{\boldsymbol \tau}_{\X}$ are linear transformations of $ \{ \hbxmq, q = 1,\dots,Q,m=1,\dots,M \} $, $\widehat { \boldsymbol \tau }_{\adjtwo} $ has the smallest asymptotic covariance among all of the considered estimators (see Theorem~\ref{thm::adj3}).}

To investigate the theoretical properties of $  \widehat { \boldsymbol \tau }_{\adjtwo} $, we need to project the potential outcomes onto the space spanned by the linear combinations of the covariates within each block,
\begin{equation*}
Y_i(q) = \bar Y_{[m]}(q) + (\boldsymbol X_i - \bar{\boldsymbol X}_{[m]})^\t \boldsymbol  \beta_{[m]} (q) + \mu_i(q), \quad i \in [m],
\end{equation*}
where $\mu_i(q)$ are (fixed) projection errors. It is easy to see that $\bar{\mu}_{[m]}(q) = 0$ for $m=1,\dots,M$. Let us denote the residuals as
$$
\hat \mu_i(q) = Y_i(q) - \hbymq - \big\{ \X_i - \hbxmq \big\} ^\T \widehat {\boldsymbol \beta}_{[m]}(q) , \quad i \in [m].
$$
\red{Let $\bmu_i=(\mu_i(1),...,\mu_i(Q))^{\T}$ and $\widehat\bmu_i=(\hat\mu_i(1),...,\hat\mu_i(Q))^{\T}$. Define $\widetilde{ \boldsymbol  V}(\bmu)$, $\boldsymbol  {V}(\bmu)$, and $\widehat { \boldsymbol V}_n ( \widehat \bmu )$ similarly to $\widetilde{ \boldsymbol  V}(\Y)$, $\boldsymbol  V (\Y)$, and $\widehat { \boldsymbol V}_n ( \Y )$ except that $\Y_i$ is replaced by $\bmu_i$ and $\widehat \bmu_i$, respectively.} 

\begin{condition}\label{cond variance error3}
The following weighted variances and covariances of the projection errors $\mu_i(q)$ tend to finite limits:
\[  \sum^M_{m=1} \pi_{[m]} \dfrac{ S^2_{[m] \bmu(q)} }{ \emq }, \quad \sum^M_{m=1} \pi_{[m]} S_{[m] \bmu(q) \bmu(q') }, \quad 1\leq q,q' \leq Q. \]
\end{condition}

\begin{condition}
\label{cond large nm}
(a) The block size $\nm \rightarrow \infty$  for $m = 1,\ldots, M$, and there exists a constant $M_{\max}$, such that $M \leq M_{\max}$; (b) The block-specific covariance matrix $\S_{[m]\X\X}$ converges to a finite, invertible matrix, and the block-specific variance, $S^2_{[m]\Y(q)} $, and covariances, $\S_{[m]\X \Y(q)}$ and $\S_{[m] \Y(q) \Y(q')}$, $1\leq q, q' \leq Q$, converge to finite limits.
\end{condition}


\begin{theorem}\label{thm::adj3}
If Conditions~\ref{cond core 0}, \blue{\ref{cond second moment} for $R_i(q)=Y_i(q), \X_i$, \ref{cond variance error3}, and \ref{cond large nm} hold}, then $ \widetilde{ \boldsymbol  V}_n (\bmu) $ and $\boldsymbol  V_n (\bmu) $ have finite limits, denoted by $ \widetilde{\boldsymbol V}( \bmu )$ and ${\boldsymbol V}( \bmu )$, respectively, $ \sqrt{n} \{  \widehat{\boldsymbol \tau}_\adjtwo -\boldsymbol\tau \}   \stackrel{d}{\longrightarrow} \mathcal{N}(  \boldsymbol 0, \widetilde{ \boldsymbol  V}( \bmu ) )$, and $ \widehat { \boldsymbol V}_n ( \hat \bmu ) \stackrel{p}{\longrightarrow} {\boldsymbol V}( \bmu)  \geq   \widetilde{\boldsymbol V}( \bmu)$. Furthermore, $ \widehat{\boldsymbol \tau}_\inter$ has the smallest asymptotic covariance among the class of estimators of the \blue{following forms:
$$
\dfrac{1}{2^{K-1}}  \sumq \dq  \summ \pim \Big[ \hbymq - \Big\{ \hbxmq - \bxm \Big\}^\t  \widetilde{\bbeta}_{[m]}(q)  \Big], \quad
\widehat{\boldsymbol \tau}_{\unadj} - \summ \pim \widetilde \bGamma_{[m]}^\T  \widehat{\boldsymbol \tau}_{\X,[m]},
$$
where $  \widetilde{\bbeta}_{[m]}(q)$ and $\widetilde \bGamma_{[m]}$ ($q=1,\dots,Q$, $m=1,
\dots,M$) are estimated adjusted coefficients} that converge in probability to finite limits. The difference between the asymptotic covariances of $  \widehat{\boldsymbol \tau}_\unadj$ and $  \widehat{\boldsymbol \tau}_\adjtwo $  is the limit of
$$
 2^{-2(K-1)} \summ \pim \Big[ \sumq \dfrac{1}{\emq} \dq \bbeta^\t_{[m]}(q) \S_{[m]\X\X} \bbeta_{[m]}(q) \dq^\t   -\Big\{ \sumq \dq \bbeta^\t_{[m]}(q) \Big\} \S_{[m]\X\X} \Big\{ \sumq \dq \bbeta^\t_{[m]}(q) \Big\}^\t \Big],
$$
and the difference between the limits of the covariance estimators, $  \widehat { \boldsymbol V}_n ( \boldsymbol Y  ) $ and $  \widehat { \boldsymbol V}_n ( \hat \bmu ) $,  is the limit of
\[   \summ \pim \sumq \dfrac{1}{\emq} \dq  \bbeta^\t_{[m]}(q) \S_{[m]\X\X} \bbeta_{[m]}(q) \dq^\t \geqslant 0. \]

\end{theorem}



According to Theorem~\ref{thm::adj3}, $ \widehat{\boldsymbol \tau}_\adjtwo$ is consistent and jointly asymptotically normal, and its asymptotic covariance is no greater than those of $  \widehat{\boldsymbol \tau}_\unadj $, $  \widehat{\boldsymbol \tau}_\adj $, $ \widehat{\boldsymbol \tau}_\cond$, \blue{and $\widehat{\boldsymbol \tau}_\condfull$}. Therefore, it is the most efficient method, at least asymptotically, to infer $\btau$ \blue{(or $\boldsymbol C \boldsymbol \tau$)}  when there are only  large blocks. \blue{This is not surprising because previous works have shown that including interactions could improve efficiency \citep[see, e.g.,][]{lin2013,lu2016covariate,lu2016randomization,Liu2019,Lei2020,Su2021}.} We can construct an asymptotically conservative confidence region for \blue{$\boldsymbol C \boldsymbol \tau$},
$$
\blue{\Big\{  \boldsymbol  \mu : \  n  ( \boldsymbol C \widehat{\boldsymbol \tau}_\inter - \boldsymbol  \mu  )^\T \big\{ \boldsymbol C  \widehat{ \boldsymbol V}_n (  \hat \bmu    ) \boldsymbol C^\T \big\} ^{-1} ( \boldsymbol C \widehat{\boldsymbol \tau}_\inter - \boldsymbol  \mu  )  \leq   \chi^2_{F_1} ( 1 - \alpha )   \Big\}.}
$$

\begin{remark}
Although $ \widehat{\boldsymbol \tau}_\adjtwo$ is optimal (asymptotically) among all of the considered estimators, evidence has suggested that this block-specific covariate adjustment method can lead to inferior performance when there exist small blocks \citep{Liu2019}.
\end{remark}

\subsection{\red{Summary of covariate-adjusted estimators}}

\red{From the asymptotic analysis above, $\widehat{\boldsymbol \tau}_\adjtwo$ is the most efficient estimator among all of the considered methods. However, $\widehat{\boldsymbol \tau}_\adjtwo$ may not be applicable or have inferior performance when there exist small blocks. In such cases, we can use $ \widehat{\boldsymbol \tau}_\unadj$, $ \widehat{\boldsymbol \tau}_\adj$,  $ \widehat{\boldsymbol \tau}_\cond$, and $\widehat{\boldsymbol \tau}_\condfull$. Compared to $ \widehat{\boldsymbol \tau}_\unadj$, $ \widehat{\boldsymbol \tau}_\adj$ increases the efficiency when the propensity scores are the same across blocks but may degrade the efficiency otherwise; $ \widehat{\boldsymbol \tau}_\cond$ increases the efficiency for estimating each  factorial effect $\tau_f$ but may degrade the efficiency for estimating $\boldsymbol C \boldsymbol \tau$ for some $\boldsymbol C$. $\widehat{\boldsymbol \tau}_\condfull$ is generally more efficient than $ \widehat{\boldsymbol \tau}_\cond$ and $\widehat{\boldsymbol \tau}_\unadj$ even when the propensity scores differ across blocks. Moreover, $\widehat{\boldsymbol \tau}_\condfull$ is asymptotically equivalent to $ \widehat{\boldsymbol \tau}_\adj$ for the case of equal propensity scores across blocks but needs to estimate more adjusted coefficients ($pQF$ versus $pQ$).  Therefore, we recommend $ \widehat{\boldsymbol \tau}_\adj$ when there exist small blocks and the propensity scores are the same across blocks, $\widehat{\boldsymbol \tau}_\condfull$ when there exist small blocks and the propensity scores differ across blocks, and $ \widehat{\boldsymbol \tau}_\inter$ when there are only  large blocks.}

\section{Simulation study}
In this section, we evaluate the finite-sample performances of  the unadjusted and four covariate-adjusted estimators with a simulation study. We consider a randomized block $2^2$ factorial experiment with $Q=4$ treatment combinations, denoted as $\{-1,-1\}$, $\{-1,+1\}$, $\{+1,-1\}$, and $\{+1,+1\}$.
\blue{Besides uniform  weighting, we  also consider a treatment combination weighted by $(-2,-1,2,1)$, which is the linear transformation of the average factorial effects, $\boldsymbol C \boldsymbol \tau$ with $\boldsymbol C =(1,0,-1/3)$. We call this transformation the general-weight effect.}
The potential outcomes are generated according to the following equations:
$$Y_{i}(\{-1,-1\})=\X_{i}^{\mathrm{T}}  \boldsymbol  \beta_{11}+\exp \left(\X_{i}^{\mathrm{T}}  \boldsymbol  \beta_{12}\right)+\varepsilon_{i}(1), \quad i=1, \cdots, n, $$
$$ Y_{i}(\{-1,+1\})=\X_{i}^{\mathrm{T}}  \boldsymbol  \beta_{21}+\exp \left(\X_{i}^{\mathrm{T}}  \boldsymbol  \beta_{22}\right)+\varepsilon_{i}(2), \quad i=1, \cdots, n,
$$
$$ Y_{i}(\{+1,-1\})= \X_{i}^{\mathrm{T}}  \boldsymbol  \beta_{31}+\exp \left(\X_{i}^{\mathrm{T}}  \boldsymbol  \beta_{32}\right)+\varepsilon_{i}(3), \quad i=1, \cdots, n,
$$
$$ Y_{i}(\{+1,+1\})=\X_{i}^{\mathrm{T}}  \boldsymbol  \beta_{41}+\exp \left(\X_{i}^{\mathrm{T}}  \boldsymbol  \beta_{42}\right)+\varepsilon_{i}(4), \quad i=1, \cdots, n,
$$
where $\varepsilon_i(q)$, $i=1,\dots,n, \ q=1,\dots,4$, are independent and identically distributed (i.i.d.) Gaussian random variables  with mean zero and variance $\sigma_q^2$. We choose  $\sigma_q^2$ such that the signal-to-noise ratio equals 10.  The $\X_i$ is a three-dimensional vector of covariates generated from a multivariate normal distribution with mean zero and covariance matrix $\Sigma$: $\Sigma_{jj} = 1$ and $\Sigma_{jl} = 0.5^{|j-l|}$, $j \not = l$, $j,l = 1, 2, 3$. For $j = 1, 2, 3$, we generate the coefficient vectors from uniform distributions:
$$\beta_{11j}\sim U(-1,1), \quad \beta_{12j}\sim   U(-0.1,0.1),$$
$$\beta_{21j}\sim \beta_{11j}+U(-1,1), \quad \beta_{22j}\sim  \beta_{12j}+  U(-0.1,0.1),$$
$$\beta_{31j}\sim \beta_{21j}+U(-1,1), \quad \beta_{32j}\sim  \beta_{22j}+ U(-0.1, 0.1),$$
$$\beta_{41j}\sim \beta_{31j}+U(-1,1), \quad \beta_{42j}\sim  \beta_{32j}+  U(-0.1, 0.1).$$

The potential outcomes and covariates are both generated once and then kept fixed. We consider three different cases of number of blocks, block sizes, and propensity scores. For each case, we conduct randomized block factorial experiments $10,000$ times to compare the performances of various methods in terms of bias, standard deviation (SD), root mean square error (RMSE), empirical coverage probability (CP), and mean confidence interval length (CI length) of the $95\%$ confidence interval (CI) for each component of factorial effects and the  general-weight effect. In addition, we construct Wald-type $95\%$ confidence regions for the joint main effects and compare their areas.

\subsection{Many small blocks} \label{sec::sim-small-blocks}
The number of blocks $M$ takes values from 20 to 100, with block size $n_{[m]}=12$. The propensity scores are set to be $e_{[m]q}=1/4$, $q=1,\dots,4$, $m = 1, \dots,M$.  The block size is too small, for the last covariate adjustment method to be applicable. Therefore, we only consider four estimators: $ \widehat { \boldsymbol \tau }_{\unadj}$, $ \widehat { \boldsymbol \tau }_{\adj}$, $ \widehat { \boldsymbol \tau }_{\cond}$, and $ \widehat { \boldsymbol \tau }_{\cond2}$.



\begin{figure}[h]
	\includegraphics[width=0.9\linewidth]{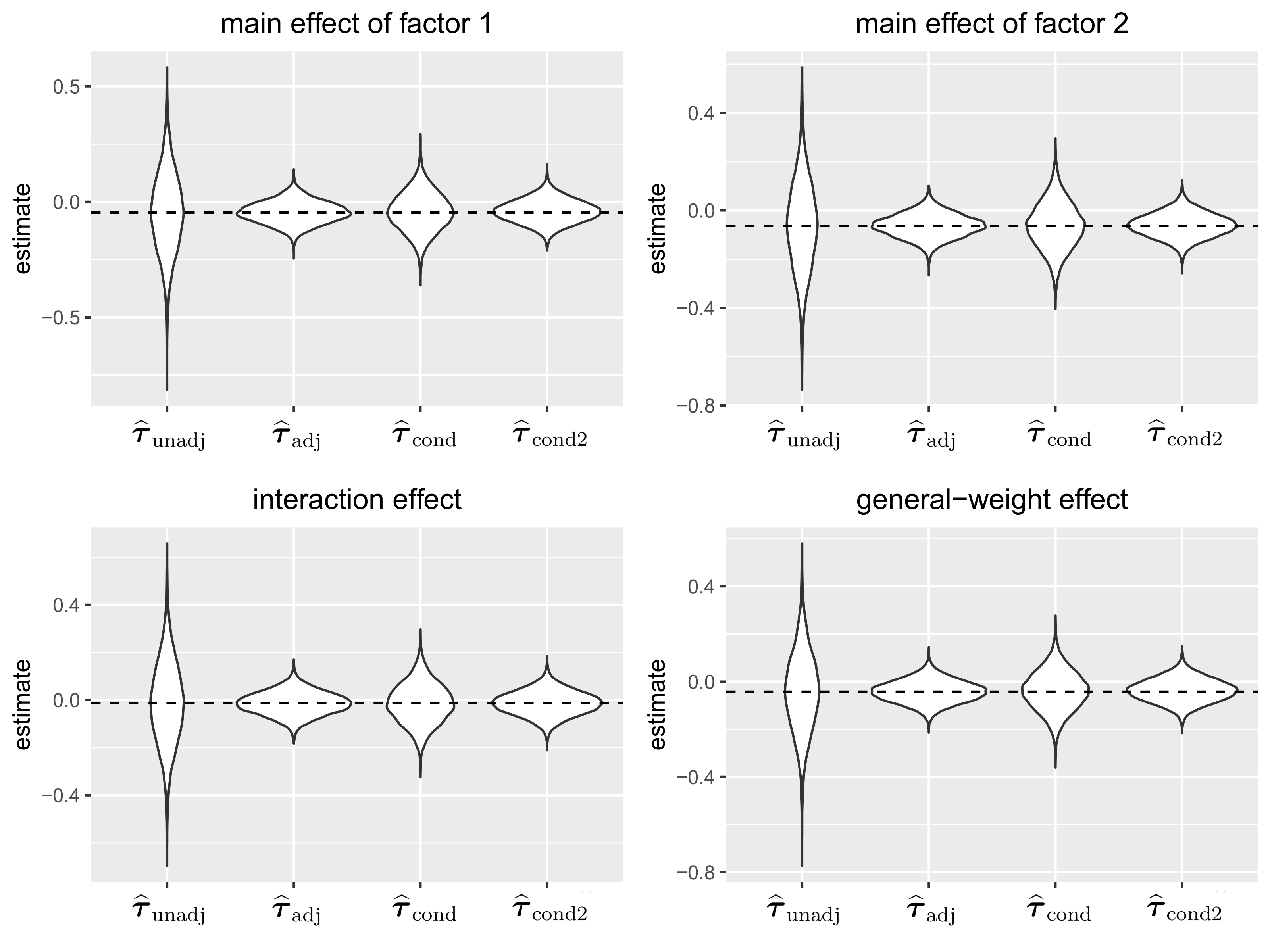}
	\caption{\label{fig::case1-violin}Violin plots of four factorial effect estimators for the case of many small blocks with $ M=20$ and $n_{[m]}=12$.}
\end{figure}

The results are shown in Tables~\ref{tab::each}--\ref{tab::joint}, Figure~\ref{fig::case1-violin}, and Figure~3 in the Supplementary Material (for RMSE and RMSE ratio). First, the biases of all methods are negligible, in accordance with the unbiasedness of $ \widehat { \boldsymbol \tau }_{\unadj}$ and the asymptotically unbiasedness of $ \widehat { \boldsymbol \tau }_{\adj}$, $ \widehat { \boldsymbol \tau }_{\cond}$, and $\widehat { \boldsymbol \tau }_{\condfull}$. Second, $ \widehat { \boldsymbol \tau }_{\adj}$, $ \widehat { \boldsymbol \tau }_{\cond}$, and $ \widehat { \boldsymbol \tau }_{\condfull}$ decrease the RMSE and thus improve the efficiency compared with $ \widehat { \boldsymbol \tau }_{\unadj}$. For example, when $M=20$ and $n_{[m]}=12$, the RMSE ratio, CI length ratio, and area ratio of confidence regions of $ \widehat { \boldsymbol \tau }_{\adj}$ relative to $ \widehat { \boldsymbol \tau }_{\unadj}$ are \blue{approximately $28\%$}, $28\%$, and $8\%$, respectively.  \blue{Third, although $ \widehat { \boldsymbol \tau }_{\adj}$ and $ \widehat { \boldsymbol \tau }_{\condfull}$ are asymptotically equivalent, $ \widehat { \boldsymbol \tau }_{\adj}$ has better finite-sample performance for the case of equal propensity scores across blocks (it is  actually the best-performing estimator in this case). This is mainly because we need to estimate fewer adjusted coefficients for $ \widehat { \boldsymbol \tau }_{\adj}$ than for $ \widehat { \boldsymbol \tau }_{\condfull}$ ($pQ$ versus $pQF$). Fourth, $ \widehat { \boldsymbol \tau }_{\cond}$ does not perform as well as $\widehat { \boldsymbol \tau }_{\condfull}$.} Finally, the percentage of improvement is almost constant as the sample size increases.



\begin{table}
\caption{\label{tab::each}Simulation results for the case of many small blocks with $M=20$ and $n_{[m]}=12$}
 \begin{threeparttable}
\begin{tabular}{ccccccccc}
		\hline
Effect & Method & Bias & SD & RMSE & RMSE ratio & CP & CI length & Length ratio \\
  \hline
main effect & $\widehat { \boldsymbol \tau }_{\unadj}$ & -0.000 & 0.171 & 0.171 & 1.000 & 0.960 & 0.714 & 1.000 \\
of factor 1 &  $\widehat { \boldsymbol \tau }_{\adj}$ & 0.001 & 0.048 & 0.048 & 0.281 & 0.964 & 0.203 & 0.284 \\
&  $\widehat { \boldsymbol \tau }_{\cond}$ & 0.003 & 0.084 & 0.084 & 0.491 & 0.981 & 0.402 & 0.563 \\
&  $\widehat { \boldsymbol \tau }_{\condfull}$ & 0.006 & 0.050 & 0.051 & 0.296 & 0.996 & 0.297 & 0.415 \\
  \hline
main effect & $\widehat { \boldsymbol \tau }_{\unadj}$ & -0.000 & 0.179 & 0.179 & 1.000 & 0.952 & 0.714 & 1.000 \\
of factor 2 &  $\widehat { \boldsymbol \tau }_{\adj}$ & 0.000 & 0.048 & 0.048 & 0.270 & 0.963 & 0.203 & 0.284 \\
&  $\widehat { \boldsymbol \tau }_{\cond}$ & 0.000 & 0.096 & 0.096 & 0.534 & 0.959 & 0.402 & 0.563 \\
&  $\widehat { \boldsymbol \tau }_{\condfull}$ & 0.003 & 0.049 & 0.049 & 0.275 & 0.987 & 0.248 & 0.347 \\
  \hline
 interaction & $\widehat { \boldsymbol \tau }_{\unadj}$ & 0.001 & 0.171 & 0.171 & 1.000 & 0.959 & 0.714 & 1.000 \\
 effect &  $\widehat { \boldsymbol \tau }_{\adj}$  & -0.000 & 0.049 & 0.049 & 0.284 & 0.963 & 0.203 & 0.284 \\
&  $\widehat { \boldsymbol \tau }_{\cond}$ & -0.001 & 0.084 & 0.084 & 0.489 & 0.983 & 0.402 & 0.563 \\
&  $\widehat { \boldsymbol \tau }_{\condfull}$ & 0.001 & 0.051 & 0.051 & 0.297 & 0.996 & 0.299 & 0.418 \\
   \hline
general-weight & $\widehat { \boldsymbol \tau }_{\unadj}$ & -0.001 & 0.163 & 0.163 & 1.000 & 0.961 & 0.682 & 1.000 \\
 effect &  $\widehat { \boldsymbol \tau }_{\adj}$  & 0.001 & 0.047 & 0.047 & 0.286 & 0.964 & 0.196 & 0.287 \\
&  $\widehat { \boldsymbol \tau }_{\cond}$ & 0.003 & 0.083 & 0.083 & 0.511 & 0.980 & 0.394 & 0.578 \\
&  $\widehat { \boldsymbol \tau }_{\condfull}$ & 0.006 & 0.049 & 0.049 & 0.301 & 0.996 & 0.287 & 0.420 \\
   \hline
  \end{tabular}
 \begin{tablenotes}
 \item Note: SD, standard deviation; RMSE, root mean squared error; RMSE ratio, ratio of RMSE relative to that of $\widehat { \boldsymbol \tau }_{\unadj}$; CP, empirical coverage probability of $95\%$ confidence interval; CI length, mean confidence interval length; Length ratio, ratio of mean confidence interval length relative to that of $\widehat { \boldsymbol \tau }_{\unadj}$.
 \end{tablenotes}
 \end{threeparttable}
\end{table}

\begin{table}
\centering
\caption{\label{tab::joint}Areas of the $95\%$ confidence regions for the joint main effects and area ratios}
\begin{tabular}{ccccc}
  \hline
method  & \multicolumn{2}{c}{ many small blocks ($M=20$)}  &  \multicolumn{2}{c}{two large heterogeneous blocks ($\nm = 108$)} \\
 & Ellipse area & Area ratio   & Ellipse area & Area ratio \\
  \hline
  $\widehat { \boldsymbol \tau }_{\unadj}$ & 0.063 & 1.000  & 0.210 & 1.000 \\
  $\widehat { \boldsymbol \tau }_{\adj}$ & 0.005 & 0.081     & 0.060 & 0.287 \\
  $\widehat { \boldsymbol \tau }_{\cond}$ & 0.019 & 0.305   & 0.080 & 0.381 \\
    $\widehat { \boldsymbol \tau }_{\condfull}$ & 0.009 & 0.136  & 0.069 & 0.329 \\
  $\widehat { \boldsymbol \tau }_{\inter}$ & - & - &  0.017 & 0.081 \\
   \hline
\end{tabular}
\end{table}

\subsection{Two large heterogeneous blocks}
We set the number of blocks $M = 2$ and change the block size $n_{[m]}$ from 60 to 156. The propensity scores are the same across blocks, $e_{[m]q}=1/4$, $q=1,\dots,4$, $m = 1, 2$. The coefficients $\beta_{q1j}$ and $\beta_{q2j}$, for $q=1,2,3,4$, are generated separately and independently for different blocks.

The results are shown in Tables~\ref{tab::joint}--\ref{tab::case3}, Figure~\ref{fig::case3-violin}, and Figure~4 in the Supplementary Material (for RMSE and RMSE ratio). We can see that $ \widehat { \boldsymbol \tau }_{\inter}$ performs the best. When $n_{[m]}=108$, $ \widehat { \boldsymbol \tau }_{\inter}$ improves the RMSE, CI length, and area of the confidence region of $ \widehat { \boldsymbol \tau }_{\adj}$ by approximately $50\%$, $50\%$, and $70\%$, respectively. Because $ \widehat { \boldsymbol \tau }_{\adj}$, $ \widehat { \boldsymbol \tau }_{\cond}$, \blue{and $ \widehat { \boldsymbol \tau }_{\condfull}$} pool the heterogeneous blocks together, they lose efficiency when compared to $ \widehat { \boldsymbol \tau }_{\inter}$.


\begin{figure}
	\includegraphics[width=0.9\linewidth]{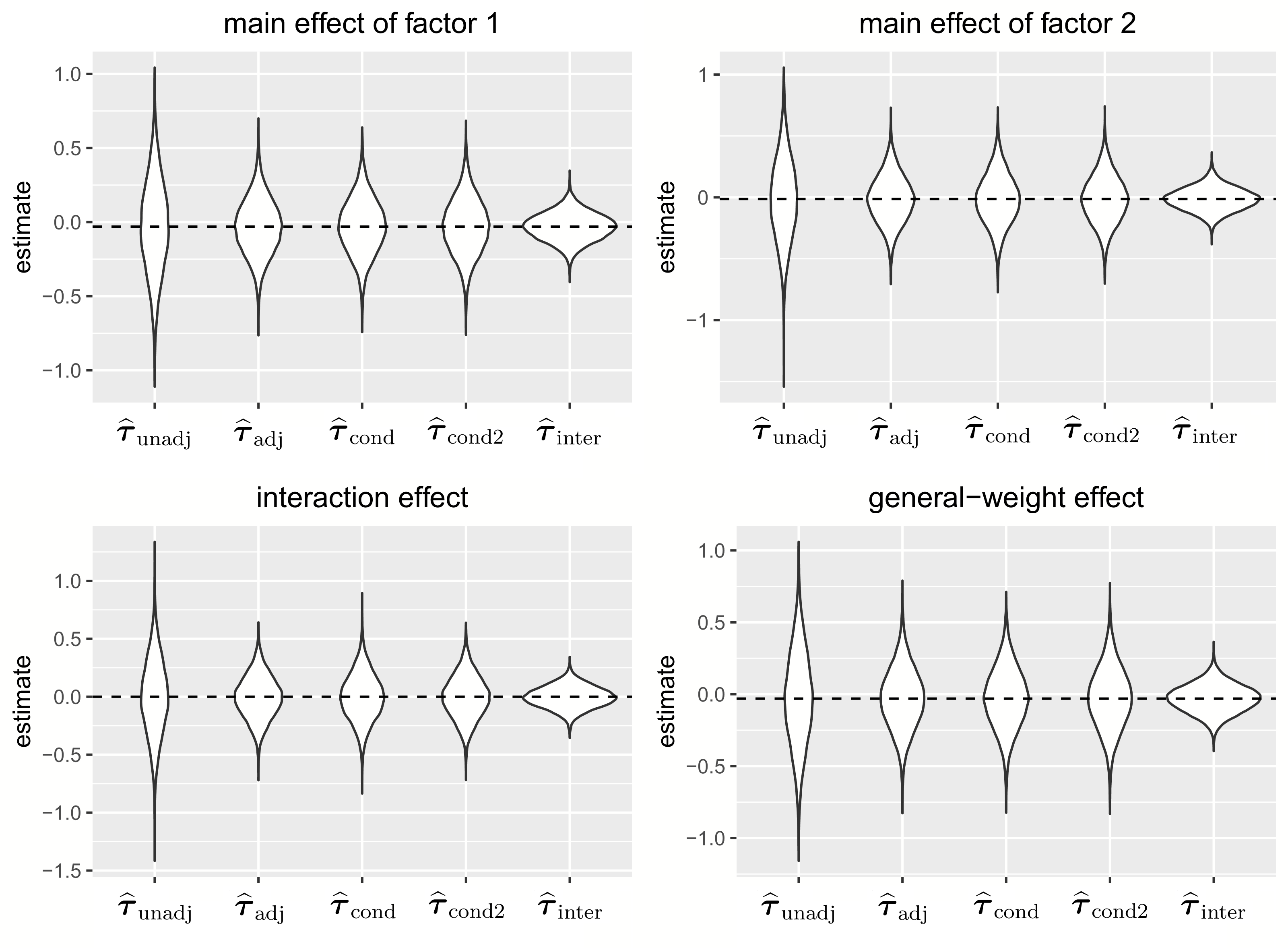}
	\caption{\label{fig::case3-violin}Violin plots of five factorial effect estimators for the case of two  large heterogeneous blocks with  $n_{[m]}=108$.}
\end{figure}

\begin{table}
	\caption{\label{tab::case3}Simulation results for the case of two large heterogeneous blocks with $n_{[m]}=108$}
	 \begin{threeparttable}
	\begin{tabular}{ccccccccc}
		\hline
		Effect & Method & Bias & SD & RMSE & RMSE ratio & CP & CI length & Length ratio \\
		\hline
main effect & $\widehat { \boldsymbol \tau }_{\unadj}$  & 0.001 & 0.313 & 0.313 & 1.000 & 0.960 & 1.304 & 1.000 \\
of factor 1 &  $\widehat { \boldsymbol \tau }_{\adj}$  & -0.002 & 0.187 & 0.187 & 0.597 & 0.947 & 0.731 & 0.561 \\
&  $\widehat { \boldsymbol \tau }_{\cond}$ & 0.002 & 0.185 & 0.185 & 0.591 & 0.970 & 0.819 & 0.628 \\
&  $\widehat { \boldsymbol \tau }_{\condfull}$ & 0.001 & 0.188 & 0.188 & 0.600 & 0.963 & 0.800 & 0.614 \\
&  $\widehat { \boldsymbol \tau }_{\inter}$  & -0.001 & 0.093 & 0.093 & 0.296 & 0.951 & 0.371 & 0.285 \\ 		\hline

main effect & $\widehat{ \boldsymbol \tau }_{\unadj}$ & -0.003 & 0.331 & 0.331 & 1.000 & 0.952 & 1.304 & 1.000 \\
of factor 2 &  $\widehat{ \boldsymbol \tau }_{\adj}$  & -0.003 & 0.191 & 0.191 & 0.577 & 0.942 & 0.731 & 0.561 \\
&  $\widehat{ \boldsymbol \tau }_{\cond}$ & -0.003 & 0.204 & 0.204 & 0.616 & 0.951 & 0.819 & 0.628 \\
&  $\widehat{ \boldsymbol \tau }_{\condfull}$ & -0.003 & 0.191 & 0.191 & 0.577 & 0.945 & 0.742 & 0.569 \\
&  $\widehat{ \boldsymbol \tau }_{\inter}$  & -0.001 & 0.093 & 0.093 & 0.281 & 0.950 & 0.371 & 0.285 \\
		\hline

interaction & $\widehat { \boldsymbol \tau }_{\unadj}$  & -0.003 & 0.324 & 0.324 & 1.000 & 0.955 & 1.304 & 1.000 \\
effect &  $\widehat { \boldsymbol \tau }_{\adj}$ & -0.005 & 0.184 & 0.184 & 0.569 & 0.950 & 0.731 & 0.561 \\
&  $\widehat { \boldsymbol \tau }_{\cond}$ & -0.003 & 0.200 & 0.200 & 0.619 & 0.955 & 0.819 & 0.628 \\
&  $\widehat { \boldsymbol \tau }_{\condfull}$ & -0.005 & 0.184 & 0.185 & 0.569 & 0.953 & 0.738 & 0.566 \\
&  $\widehat { \boldsymbol \tau }_{\inter}$  & -0.004 & 0.093 & 0.094 & 0.289 & 0.953 & 0.371 & 0.285 \\
		\hline

general-weight & $\widehat { \boldsymbol \tau }_{\unadj}$  & 0.002 & 0.331 & 0.331 & 1.000 & 0.960 & 1.373 & 1.000 \\
effect &  $\widehat { \boldsymbol \tau }_{\adj}$ & -0.000 & 0.211 & 0.211 & 0.637 & 0.944 & 0.819 & 0.597 \\
&  $\widehat { \boldsymbol \tau }_{\cond}$ & 0.002 & 0.210 & 0.210 & 0.635 & 0.964 & 0.910 & 0.663 \\
&  $\widehat { \boldsymbol \tau }_{\condfull}$ & 0.003 & 0.212 & 0.212 & 0.640 & 0.959 & 0.884 & 0.644 \\
&  $\widehat { \boldsymbol \tau }_{\inter}$  & 0.000 & 0.097 & 0.097 & 0.292 & 0.949 & 0.385 & 0.280 \\
		\hline
	\end{tabular}
	 \begin{tablenotes}
 \item
Note: SD, standard deviation; RMSE, root mean squared error; RMSE ratio, ratio of RMSE relative to that of $\widehat { \boldsymbol \tau }_{\unadj}$; CP, empirical coverage probability of $95\%$ confidence interval; CI length, mean confidence interval length; Length ratio, ratio of mean confidence interval length relative to that of $\widehat { \boldsymbol \tau }_{\unadj}$.
 \end{tablenotes}
 \end{threeparttable}
\end{table}

\subsection{An example with unequal propensity scores}
In this section, we provide an example to show that $\widehat { \boldsymbol \tau }_{\adj}$ may lose efficiency when the propensity scores differ across blocks, while $\widehat { \boldsymbol \tau }_{\cond}$ and $\widehat { \boldsymbol \tau }_{\condfull}$ do not. We consider $K=2$ factors and set number of blocks $M = 10$ with block size $n_{[m]}=40$. The  propensity scores are
$$\boldsymbol{e}_{[m]}=\Big(  \frac{m}{2M} , \frac{m}{2M}, 0.5 - \frac{m}{2M} , 0.5 - \frac{m}{2M} \Big)^\T, \quad m = 1, \dots, 5,$$
$$\boldsymbol{e}_{[m]}= \Big(  0.5 - \frac{m - 5}{2M} , 0.5 - \frac{m - 5}{2M} , \frac{m - 5}{2M} , \frac{m - 5}{2M}  \Big)^\T, \quad m = 6, \dots, 10. $$
The potential outcomes are generated as follows:
$$
Y_{i}(\{-1,-1\})=-10e_{[m]1}X_{i}+\varepsilon_{i}(1), \quad i \in [m],
$$
$$
Y_{i}(\{-1,+1\})=-10e_{[m]2}X_{i}+\varepsilon_{i}(2),\quad i\in [m],
$$
$$
Y_{i}(\{+1,-1\})=10e_{[m]3}\exp\{e_{[m]3}X_{i}\}+\varepsilon_{i}(3), \quad i\in [m],
$$
$$
Y_{i}(\{+1,+1\})=10e_{[m]4}\exp\{e_{[m]4}X_{i}\}+\varepsilon_{i}(4), \quad i\in [m],
$$
where $\varepsilon_i(1)$, $\varepsilon_i(2)$, $\varepsilon_i(3)$, $\varepsilon_i(4)$ are generated from Gaussian distribution with mean zero and variance 0.01. The $X_i$ is an one-dimensional covariate generated from a standard normal distribution. In this case, $\widehat { \boldsymbol \tau }_{\inter}$ is not applicable because some blocks are too small.

\begin{figure}
\centering
	\includegraphics[width=0.65\linewidth]{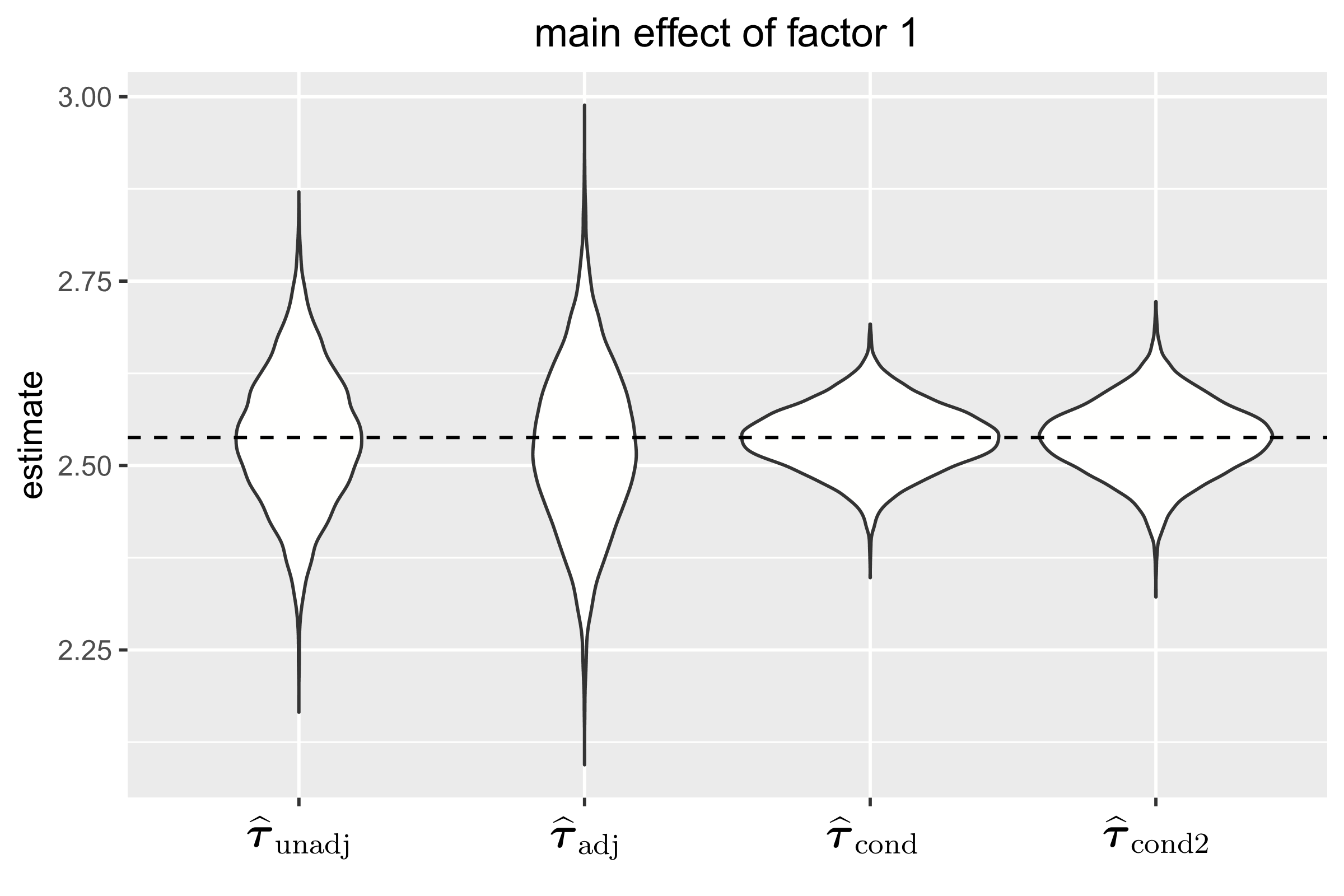}
	\caption{\label{fig:case4}Violin plots of factorial effect estimators for the example of unequal propensity scores.}
\end{figure}

The results for the main effect of factor 1 are shown in Figure~\ref{fig:case4}. It is easy to see that $ \widehat { \boldsymbol \tau }_{\adj}$ performs worse than $ \widehat { \boldsymbol \tau }_{\unadj}$. In fact, the RMSE of $ \widehat { \boldsymbol \tau }_{\adj}$ is $119.4\%$ of that of $ \widehat { \boldsymbol \tau }_{\unadj}$. In contrast, $ \widehat { \boldsymbol \tau }_{\cond}$ \blue{and $ \widehat { \boldsymbol \tau }_{\condfull}$} still perform well, reducing the RMSE of $ \widehat { \boldsymbol \tau }_{\unadj}$ by approximately $50\%$.

\section{Application}
In this section, we analyze a real dataset from a clinical trial, CALGB 40603, using the proposed methods. CALGB 40603 was a randomized block  $2^2$ factorial  phase II trial, that sought to evaluate the impact of adding bevacizumab and/or carboplatin on pathologic complete response (pCR) rates in patients with Stage II to III triple-negative breast cancer (TNBC) \citep{Sikov2015}.
For standard neoadjuvant chemotherapy, patients with TNBC received paclitaxel 80 mg/$m^2$ once per week for 12 weeks, followed by doxorubicin plus cyclophosphamide once every 2 weeks for four cycles. Factor 1 was adding bevacizumab (10 mg/kg once every 2 weeks for nine cycles), and factor 2 was adding carboplatin (once every 3 weeks for four cycles) to the standard neoadjuvant chemotherapy. The 443 patients were blocked by pretreatment clinical stage (II or III) and randomly assigned into four treatment arms with equal probabilities:
\begin{itemize}
\item Arm C: standard neoadjuvant chemotherapy,

\item Arm A: standard neoadjuvant chemotherapy + bevacizumab,

\item Arm B: standard neoadjuvant chemotherapy + carboplatin,

\item Arm AB: standard neoadjuvant chemotherapy + bevacizumab + carboplatin.
\end{itemize}

The outcome of interest is the pCR breast, defined as the absence of residual invasive disease with or without ductal carcinoma in situ (ypT0/is). Removal of the patients with missing outcomes leaves 433 patients, 295 in clinical stage II and 138 in clinical stage III. We consider eight baseline covariates for adjustments, including tumor grade, clinical T stage, clinical N stage, and so on.


\begin{table}
\centering
\caption{\label{tab::real}Point estimators and $95\%$ confidence intervals for the factorial effects of adding bevacizumab or/and carboplatin, and the reduction of variance relative to the unadjusted estimator}
 \begin{threeparttable}
\begin{tabular}{ccccc}
\hline
Method &main effect of  & main effect of   &   interaction effect of  & reduction of variance\\
& bev & carbo  &   bev and carbo & \\
\hline

$\widehat { \boldsymbol \tau }_{\unadj}$ & 0.140 &0.113 &0.029 &0  \\

& [0.047, 0.233] &[0.020, 0.207] &  [-0.064, 0.122]&\\

$\widehat { \boldsymbol \tau }_{\adj}$  & 0.159 & 0.110 & 0.017 &12.8\% \\

&[0.072, 0.246] & [0.023, 0.197] & [-0.070,0.104] &\\

$\widehat { \boldsymbol \tau }_{\cond}$ & 0.168  & 0.122 &0.020 &4.9\%  \\

& [0.078, 0.259] & [0.032, 0.213]&[-0.071, 0.111]&\\

$\widehat { \boldsymbol \tau }_{\condfull}$ & 0.162  & 0.112 &0.011 &11.6\%  \\

& [0.074, 0.250] & [0.023, 0.201]&[-0.078, 0.099]&\\

$\widehat { \boldsymbol \tau }_{\inter}$ & 0.140 &0.124 & 0.018  & 21.2\%  \\
& [0.057, 0.222] & [0.041, 0.206]& [-0.065, 0.101] &\\
\hline
\end{tabular}
 \begin{tablenotes}
 \item Note: bev, bevacizumab; carbo, carboplatin.
 \end{tablenotes}
 \end{threeparttable}
\end{table}

\begin{figure}[ht]
\centering
	\includegraphics[width=0.85\linewidth]{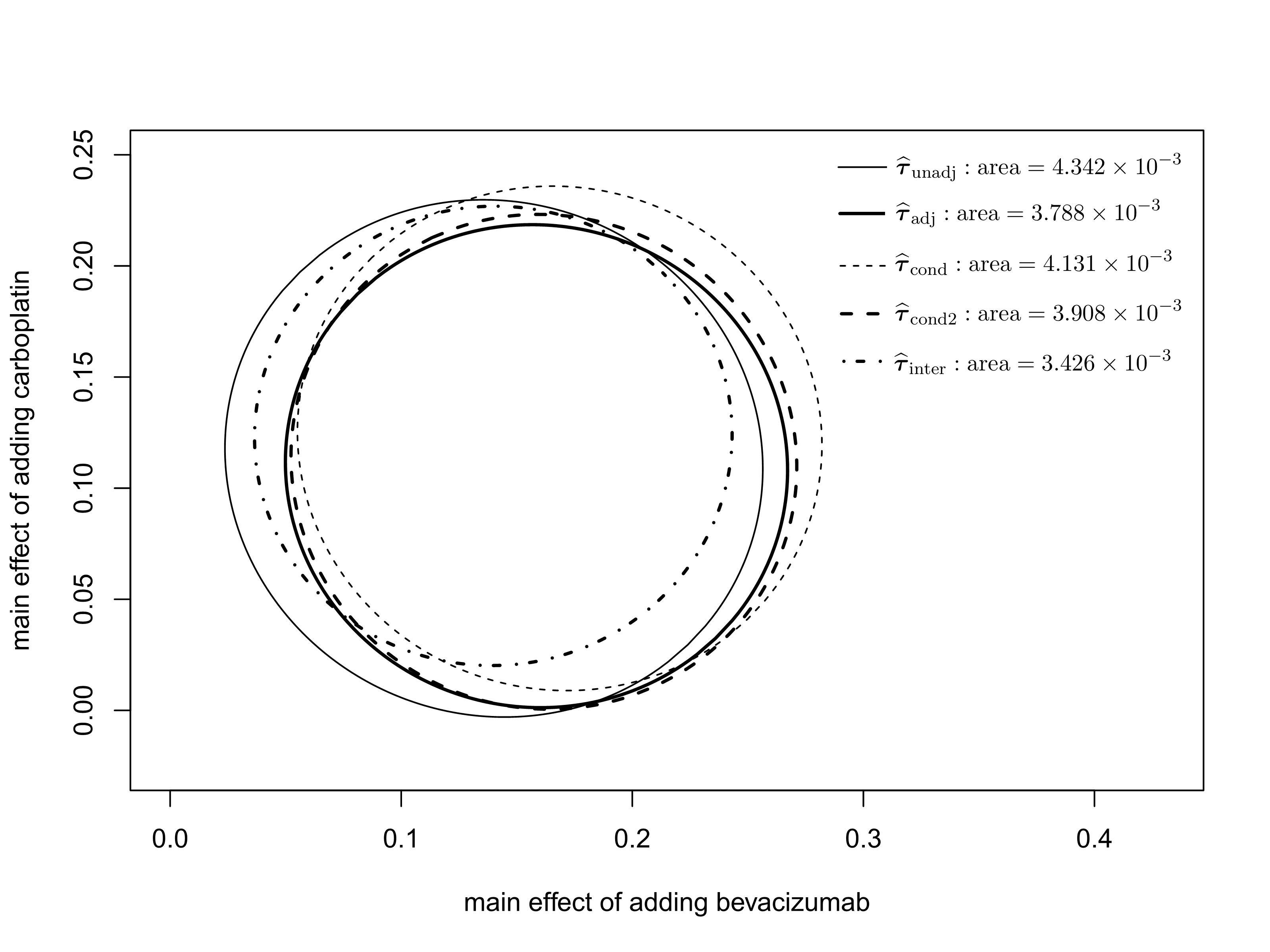}
	\caption{\label{fig::real}$95\%$ confidence region for main effects of adding bevacizumab and carboplatin.}
\end{figure}

The point estimators and $95\%$ CIs for each factorial effect are given in Table~\ref{tab::real}. Based on $\widehat { \boldsymbol \tau }_{\unadj}$, adding bevacizumab improves the pCR rate by approximately $15\%$; adding carboplatin improves the pCR rate by approximately $11\%$; and no significant interaction effect is found for adding bevacizumab and carboplatin. These conclusions are in accordance with those obtained by \cite{Sikov2015}. The covariate adjustment methods, $ \widehat { \boldsymbol \tau }_{\adj}$, $ \widehat { \boldsymbol \tau }_{\cond}$, \blue{$ \widehat { \boldsymbol \tau }_{\condfull}$}, and $ \widehat { \boldsymbol \tau }_{\inter}$, give similar statistical conclusions. However, it is interesting to note that these four methods improve efficiency, as they reduce the variance by $12.8\%$, $4.9\%$, \blue{$11.6\%$}, and $21.2\%$, respectively. Because the blocks are large and likely heterogeneous, $ \widehat { \boldsymbol \tau }_{\inter}$ performs the best. In addition, we construct Wald-type $95\%$ confidence regions for the joint main effects, which are shown in Figure~\ref{fig::real}. Compared with $\widehat { \boldsymbol \tau }_{\unadj}$, the covariate adjustment methods, $ \widehat { \boldsymbol \tau }_{\adj}$, $ \widehat { \boldsymbol \tau }_{\cond}$, \blue{$\widehat { \boldsymbol \tau }_{\condfull}$}, and $ \widehat { \boldsymbol \tau }_{\inter}$ reduce the areas of the confidence regions by $12.8\%$, $4.9\%$, \blue{$10.0\%$}, and $21.1\%$, respectively.

\section{Discussion}
In this paper, we established a general finite population vector CLT to derive the joint asymptotic distribution of blocked sample means in randomized block experiments with vector outcomes and multiple treatments. This new CLT plays a crucial role in randomization-based causal inference for the average factorial effects in randomized block $2^K$ factorial experiments. Based on the CLT, we showed that the usual (unadjusted) average factorial effects estimator is consistent and asymptotically normal, without imposing strong modeling assumptions on the potential outcomes. We proposed four covariate adjustment methods to improve the estimation and inference efficiency. We derived their asymptotic distributions, proposed conservative covariance estimators, and compared their efficiencies with that of the unadjusted estimator. \red{Our results are} robust to model misspecification and can be easily extended to more general randomized block factorial experiments with multiple-level factors, $3^K$, $4^K$, and so on.

In practice, a combination of  large and small blocks may exist. In such cases, it might be more efficient to pool together small blocks into large blocks, and then use $ \widehat{\boldsymbol \tau}_\inter$. It is worth further investigating how to efficiently perform the pooling and the follow-up covariate adjustment. Moreover, in this paper, we focused on using covariate adjustment in the analysis stage to improve the estimation and inference efficiency. Covariate adjustment can also be used in the design stage, such as rerandomization  \citep{Morgan2012, morgan2015rerandomization, Li2018}. \blue{\cite{Branson2016} proposed a rerandomization procedure in completely randomized $2^K$ factorial experiments and \cite{Li2020factorial} established its asymptotic theory.} It would be interesting to generalize the results to randomized block $2^K$ factorial experiments. Our new CLT has already established a theoretical basis for deriving the corresponding asymptotic theory. In addition, we assume that the number of covariates is fixed. In practice, however, the number of covariates can be large, even larger than the sample size. It would also be  interesting to investigate robust and efficient covariate adjustment methods in randomized block factorial experiments with high-dimensional covariates.

\section*{Acknowledgment}
\red{The authors are grateful to the associate editor and two referees for their valuable comments. This publication is based on research using information obtained from data.projectdatasphere.org, which is maintained by Project Data Sphere. Neither Project Data Sphere nor the owner(s) of any information from the website have contributed to, approved, or are in any way responsible for the contents of this publication.}

\bigskip
\begin{center}
{\large\bf SUPPLEMENTARY MATERIAL}
\end{center}

\begin{description}

\item The supplementary material provides the proofs  and additional simulation results.

\end{description}

\bibliographystyle{agsm}

\bibliography{causal}
\end{document}